\documentclass[bibyear]{aa} 
\usepackage[varg]{txfonts}
\usepackage{amsmath}
\usepackage{amssymb}
\usepackage{graphicx}

\usepackage{float}
\usepackage{multirow}
\usepackage{booktabs}
\usepackage[usenames, dvipsnames]{color}
\usepackage{placeins}

\usepackage{bookmark}
\usepackage{natbib}
\usepackage[acronym]{glossaries}
\usepackage{hyperref}
\hypersetup{
  colorlinks = true, 
  urlcolor  = blue, 
  linkcolor = red,  
  citecolor = blue  
}
\bibpunct{(}{)}{;}{a}{}{,} 
\title{
  The VLA-COSMOS 3 GHz Large Project: 
  Average radio spectral energy distribution of active galactic nuclei}
\titlerunning{Radio SED of AGN}
\author{
     K. \Tisanic  \inst{1}  \thanks{\emph{ktisanic@phy.hr}}
  \and V. \Smolcic  \inst{1}  
  \and M. Imbrišak  \inst{1}  
  \and M. Bondi    \inst{2}
  \and G. Zamorani  \inst{2}
  \and L. Ceraj    \inst{1}
  \and E. Vardoulaki \inst{3}
  \and J. Delhaize \inst{4}
}

\institute{
    Department of Physics, 
    Faculty of Science, 
    University of Zagreb, 
    Bijeni\v{c}ka cesta 32, 
    10000 Zagreb, 
    Croatia
    \and INAF - Osservatorio di Astrofisica e Scienza dello Spazio, via Gobetti 93/3, 40129, Bologna, Italy
    \and Max-Planck-Institut für Radioastronomie, Auf dem Hügel 69, D-53121 Bonn, Germany
    \and Department of Astronomy, University of Cape Town, Private Bag X3, Rondebosch 7701, South Africa
}
\usepackage{xspace}
\newcommand{\Smolcic}{Smol\v{c}i\'{c}\xspace}
\newcommand{\Tisanic}{Tisani\'{c}\xspace}
\newcommand{\dex}{\,\mathrm{dex}}

\newcommand{\MHz}{\,\mathrm{MHz}}
\newcommand{\GHz}{\,\mathrm{GHz}}

\newcommand{\kpc}{\,\mathrm{kpc}}

\newcommand{\hr}{\,\mathrm{hours}}

\newcommand{\Jy}{\,\mathrm{Jy}}

\newcommand{\muJy}{\,\mathrm{\mu Jy}}

\newcommand{\percent}{\%}

\newacronym{AGN}{AGN}{active galactic nuclei}
\newacronym{PL}{PL}{power-law}
\newacronym{BPL}{BPL}{broken power-law}
\newacronym{RxAGN}{RxAGN}{radio-excess active galactic nuclei}
\newacronym{SED}{SED}{spectral energy distribution}
\newacronym{COSMOS}{COSMOS}{Cosmological Evolution Survey}
\newacronym{GMRT}{GMRT}{Giant Meterwave Radio Telescope }
\newacronym{VLA}{VLA}{Karl G. Jansky Very Large Array }
\newacronym{RMS}{RMS}{root-mean-square}
\newacronym{FIM}{FIM}{ Fisher information metric }
\newacronym{SKADS}{SKADS}{}
\newglossaryentry{VLA3LP}
{
  name={VLA3LP},
  description={VLA-COSMOS $3\GHz$ Large Project},
  first={VLA-COSMOS $3\GHz$ Large Project \citep[hereafter VLA3LP,][]{Smolcic:17a}},
  long={VLA-COSMOS $3\GHz$ Large Project}
}
\newglossaryentry{SKA}
{
  name={SKA},
  description={Square Kilometer Array},
  first={Square Kilometer Array \citep[SKA, ][]{Braun1996TheInterferometer,Dewdey09}},
  long={Square Kilometer Array}
}
\newglossaryentry{VLA1.4JP}
{
  name={VLA1.4JP},
  description={VLA-COSMOS $1.4\GHz$ Joint Project},
  first={VLA-COSMOS $1.4\GHz$ Joint Project \citep[hereafter VLA1.4JP,][]{Schinnerer10}},
  long={VLA-COSMOS $1.4\GHz$ Joint Project}
} 
\newglossaryentry{VLA1.4LP}
{
  name={VLA1.4LP},
  description={VLA-COSMOS $1.4\GHz$ Large Project},
  first={VLA-COSMOS $1.4\GHz$ Large Project \citep[][]{Schinnerer07}},
  long={VLA-COSMOS $1.4\GHz$ Large Project}
}
\newglossaryentry{VLA1.4DP}
{
  name={VLA1.4DP},
  description={VLA-COSMOS $1.4\GHz$ Deep Project},
  first={VLA-COSMOS $1.4\GHz$ Deep Project \citep[][]{Schinnerer10}},
  long={VLA-COSMOS $1.4\GHz$ Deep Project}
}
\newglossaryentry{MWACS}
{
  name={MWACS},
  description={The Murchison Widefield Array Commissioning Survey},
  first={the Murchison Widefield Array Commissioning Survey \citep[][]{Hurley-Walker2014TheDegrees}},
  long={The Murchison Widefield Array Commissioning Survey}
}
\newglossaryentry{LOFAR}
{
  name={LOFAR},
  description={The Low-Frequency Array},
  first={the Low-Frequency Array (LOFAR)},
  long={The LOw-Frequency ARray}
}
\newacronym{COSMOS2015}{COSMOS2015}{COSMOS2015 \citep{Laigle16}}
\newacronym{i-band}{i-band}{i-band \citep{Capak07}}
\newacronym{IRAC}{IRAC}{IRAC \citep{Sanders07}}
\newglossaryentry{MBAM}
{
  name={MBAM},
  description={Manifold Boundary approximation method},
  first={Manifold Boundary approximation method (MBAM)},
  long={Manifold Boundary approximation method},
  short={MBAM}
}

\newglossaryentry{SPAM}
{
  name={SPAM},
  description={Source Peeling and Atmospheric Modeling pipeline},
  first={Source Peeling and Atmospheric Modeling pipeline \citep[SPAM][]{Intema17} },
  long={Source Peeling and Atmospheric Modeling}
}
\newglossaryentry{ANOVA}
{
  name={ANOVA},
  description={Analysis of variance},
  first={Analysis of variance (ANOVA)},
  long={Analysis of variance}
}
\newglossaryentry{MSMF}
{
  name={MSMF},
  description={multi-scale multi-frequency synthesis algorithm},
  first={multi-scale multi-frequency synthesis algorithm \citep[hereafter MSMF,][]{Rau11} },
  long={multi-scale multi-frequency synthesis algorithm}
}
\newglossaryentry{SA}
{
  name={SA},
  description={synchrotron aging},
  first={synchrotron aging {(SA)}},
  long={synchrotron aging}
}
\newglossaryentry{SSA}
{
  name={SSA},
  description={synchrotron self-absorption},
  first={synchrotron self-absorption {(SSA)}},
  long={synchrotron self-absorption}
}
\newacronym{MCMC}{MCMC}{Markov Chain Monte Carlo}
\newacronym{PSF}{PSF}{point spread function}
\newacronym{HLAGN}{HLAGN}{moderate-to-high radiative luminosity AGN}
\newacronym{HERG}{HERG}{high-excitation radio galaxy}
\newacronym{LERG}{LERG}{low-excitation radio galaxy}
\newacronym{MLAGN}{MLAGN}{low-to-moderate radiative luminosity AGN}
\newacronym{RxHLAGN}{RxHLAGN}{moderate-to-high luminosity RxAGN}
\newacronym{RxQMLAGN}{RxQMLAGN}{low-to-moderate luminosity quiescent RxAGN}
\newacronym{RxSMLAGN}{RxSMLAGN}{low-to-moderate luminosity star-forming RxAGN}

\newcommand{\PLSpectralIndex}{$0.64\pm0.07$\xspace}
\newcommand{\BPLSpectralIndexLower}{$\alpha_1=0.28\pm 0.03$\xspace}
\newcommand{\BPLSpectralIndexHigher}{$\alpha_2= 1.16\pm0.04$\xspace}
\newcommand{\BPLBreakFrequency}{$\nu_b=(4.1\pm0.2)\GHz$\xspace}
\newcommand{\BPLBreakFixedFrequency}{$4\GHz$\xspace}
\newcommand{\SSASASpectralIndexSSA}{$\alpha_{SSA}=0.8\pm0.6$\xspace}
\newcommand{\SSASASpectralIndexSA}{$\alpha_{SA}=0.4\pm0.5$\xspace}
\newcommand{\SSASAFraction}{$f=0.7\pm0.3$\xspace}
\newcommand{\SSASABreakFrequencySSA}{$\nu_1=(1\pm 2)\GHz$\xspace}
\newcommand{\SSASABreakFrequencySA}{$\nu_b=(6\pm 2)\GHz$\xspace}
\newcommand{\SSASASpectralBreak}{$\Delta\alpha=2\pm1$\xspace}
\newcommand{\SSASASpectralIndexSSAMBAM}{$\alpha_{SSA}=1.1\pm0.5$\xspace}
\newcommand{\SSASASpectralIndexSAMBAM}{$\alpha_{SA}=0.7\pm0.2$\xspace}
\newcommand{\SSASABreakFrequencySSAMBAM}{$\nu_1=(3.0\pm0.3)\GHz$\xspace}
\newcommand{\SSASAFractionMBAM}{$f=0.8\pm0.3$\xspace}
\newcommand{\flatPL}{$\alpha=0.41\pm0.07$\xspace}
\newcommand{\flatBPLlow}{$\alpha_1=0.1\pm0.1$\xspace}
\newcommand{\flatBPLhigh}{$\alpha_2=0.55\pm0.09$\xspace}
\newcommand{\flatBPLFrequency}{$\nu_b=(2.7\pm0.5)\GHz$\xspace}
\newcommand{\flatSSA}{$\alpha_{SSA}=0.53\pm0.06$\xspace}
\newcommand{\flatSSAFreq}{$\nu_1=(0.9\pm 0.1)\GHz$\xspace}
\newcommand{\FigMainSEDs}{
\begin{figure*}
\includegraphics[width=\columnwidth]{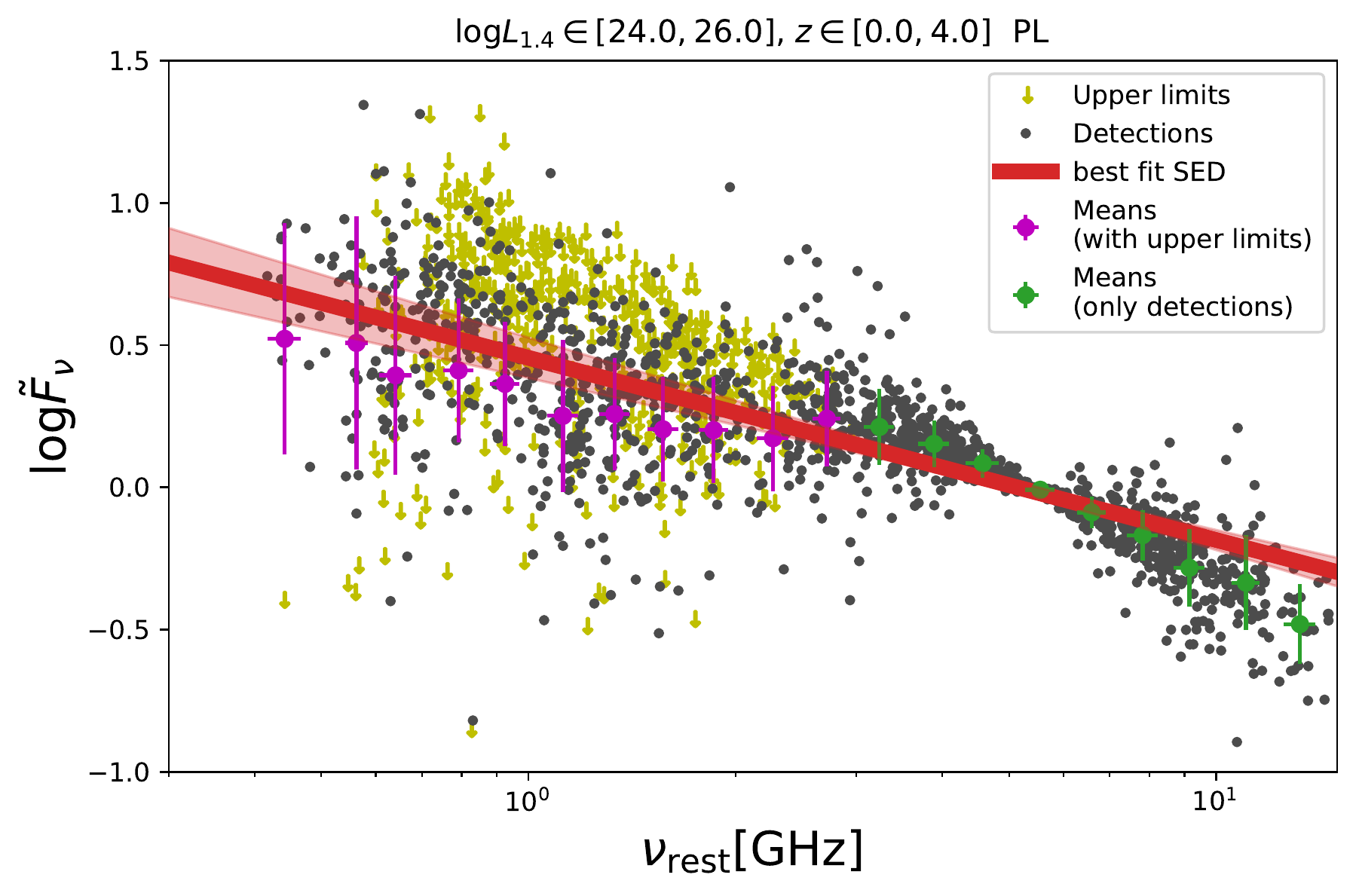}
\includegraphics[width=\columnwidth]{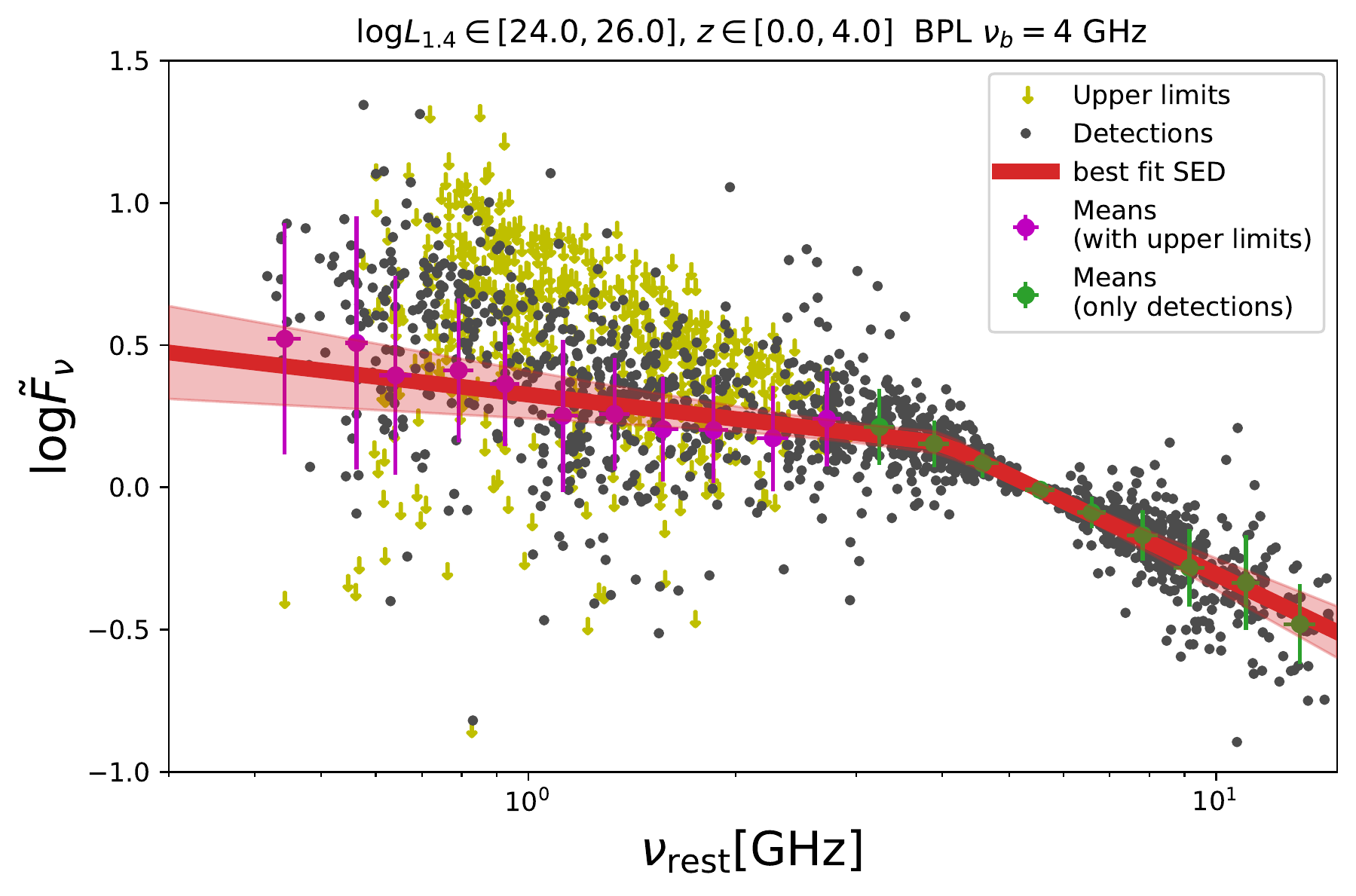}
\caption{{Radio} \acrshort{SED} of \acrshort{RxAGN} as normalized log-flux {density} vs. rest-frame frequency for $z\in[0,4]$ and $\log L_{1.4}\,\mathrm{{[W/Hz]}}\in[24,26]$. {A single \acrlong{PL} (left panel) and a \acrlong{BPL} (right panel) model of the radio \gls{SED} are shown as red lines with red confidence intervals.} Detections are shown as {gray} points, upper limits as yellow arrows, bins without upper limits as green points and bins with upper limits as magenta points. { The plots show the best fitting \acrlong{PL} (\PLSpectralIndex) and \acrlong{BPL} (\BPLSpectralIndexLower, \BPLSpectralIndexHigher and \BPLBreakFrequency) models.}}\label{fig:PLSED}
\end{figure*}}
\newcommand{\FigMBAMa}{\begin{figure}[H]
  \centering
  \includegraphics[width=.85\columnwidth]{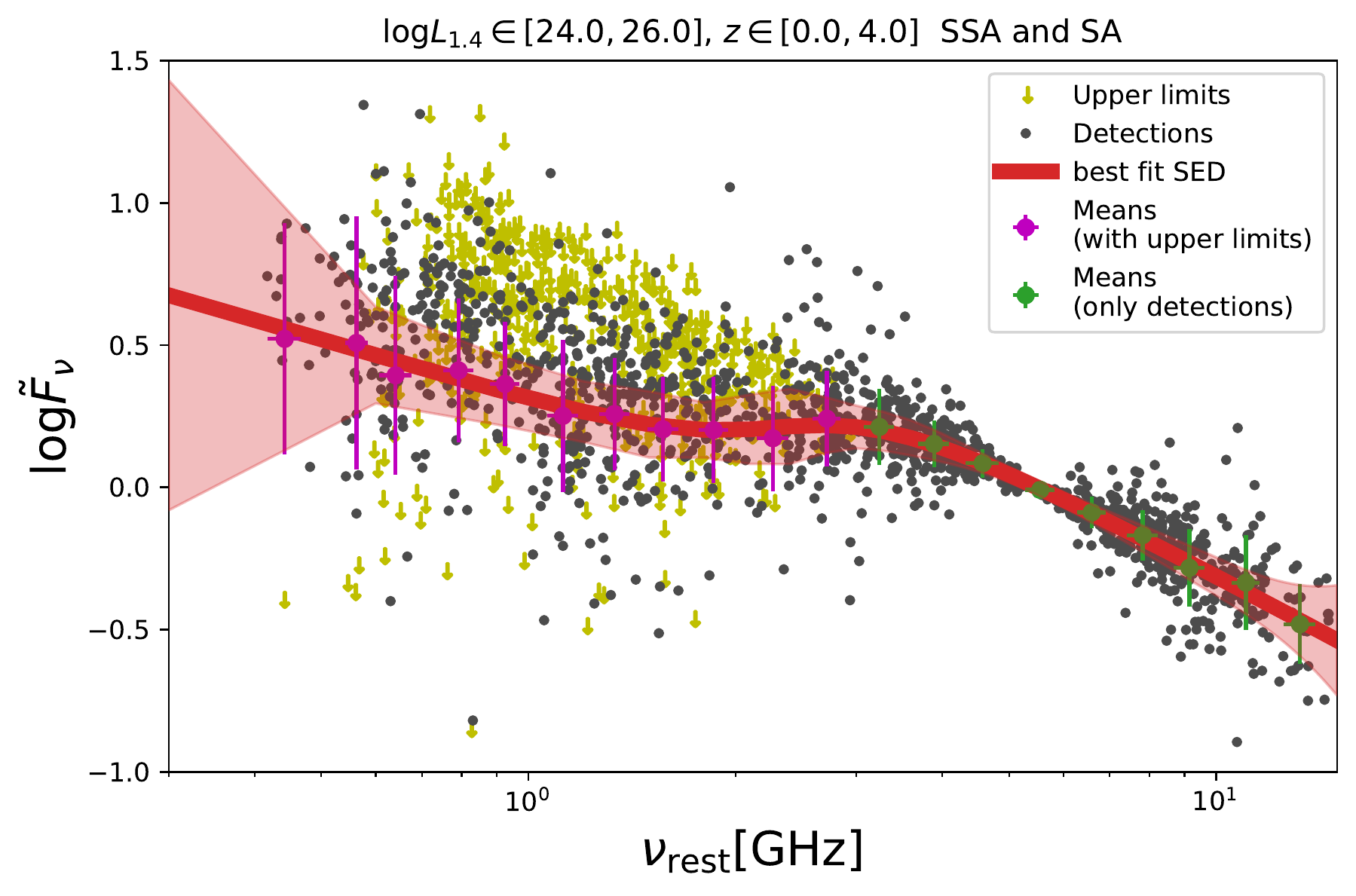}
  
   \includegraphics[width=.85\columnwidth]{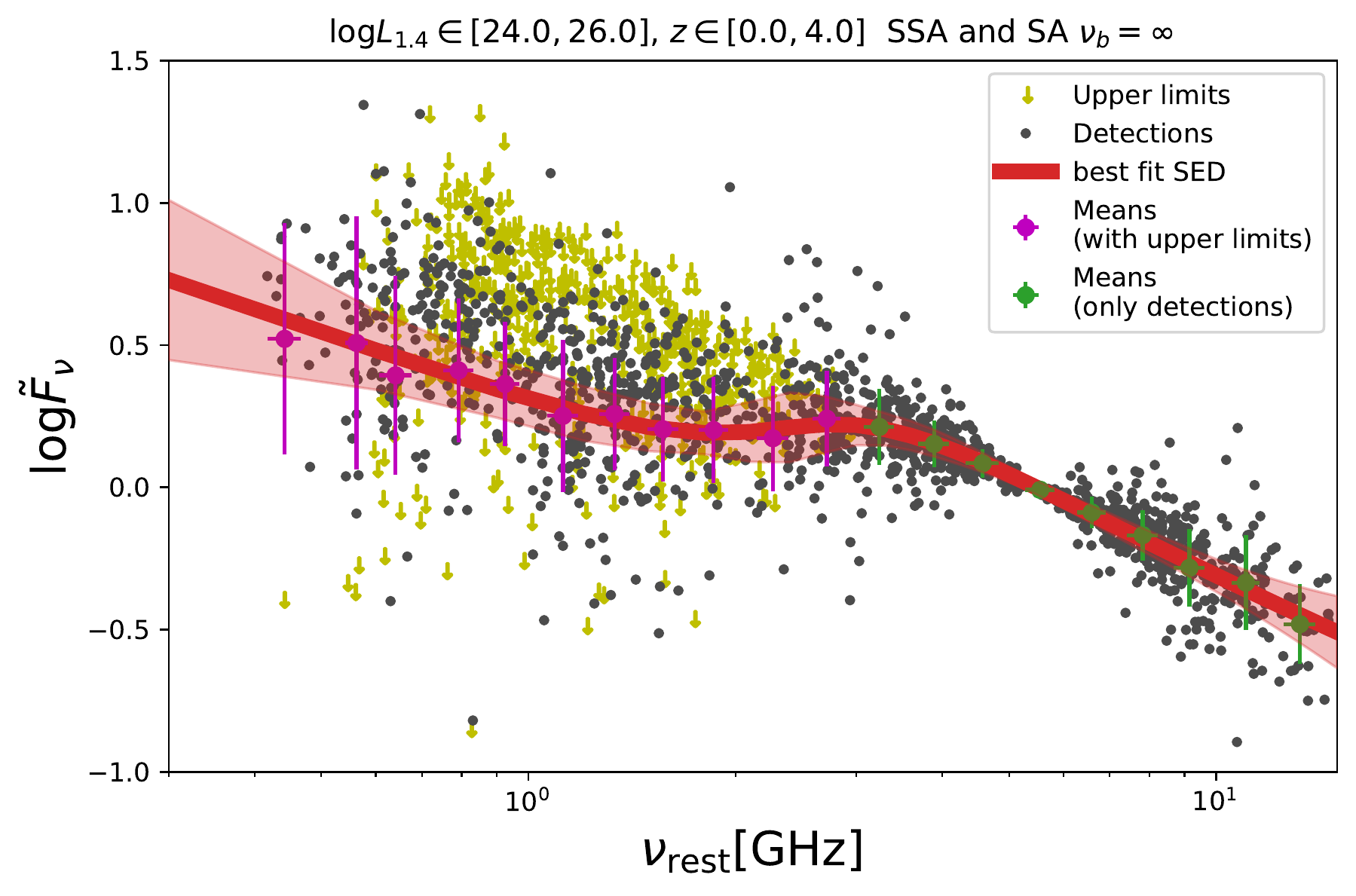}
  \caption{Average {radio} \acrshort{SED} of the \acrshort{RxAGN} sample with the model from Eq. \ref{eq:SSA+SA} {fit}. {Upper panel shows the resulting model before applying the {\acrshort{MBAM} model reduction method (described in Sect. \ref{sect:MethodsMBAM}}), while the lower panel shows the best-fitting model after reducing the number of parameters with the \acrshort{MBAM} method, {used to better constrain this complex model.}
  }}\label{fig:MBAM1}
\end{figure}}
\newcommand{\FigMBAMb}{\begin{figure*}
  \centering
  \includegraphics[width=1.1\textwidth,trim=3cm 0cm 0cm 0cm,clip]{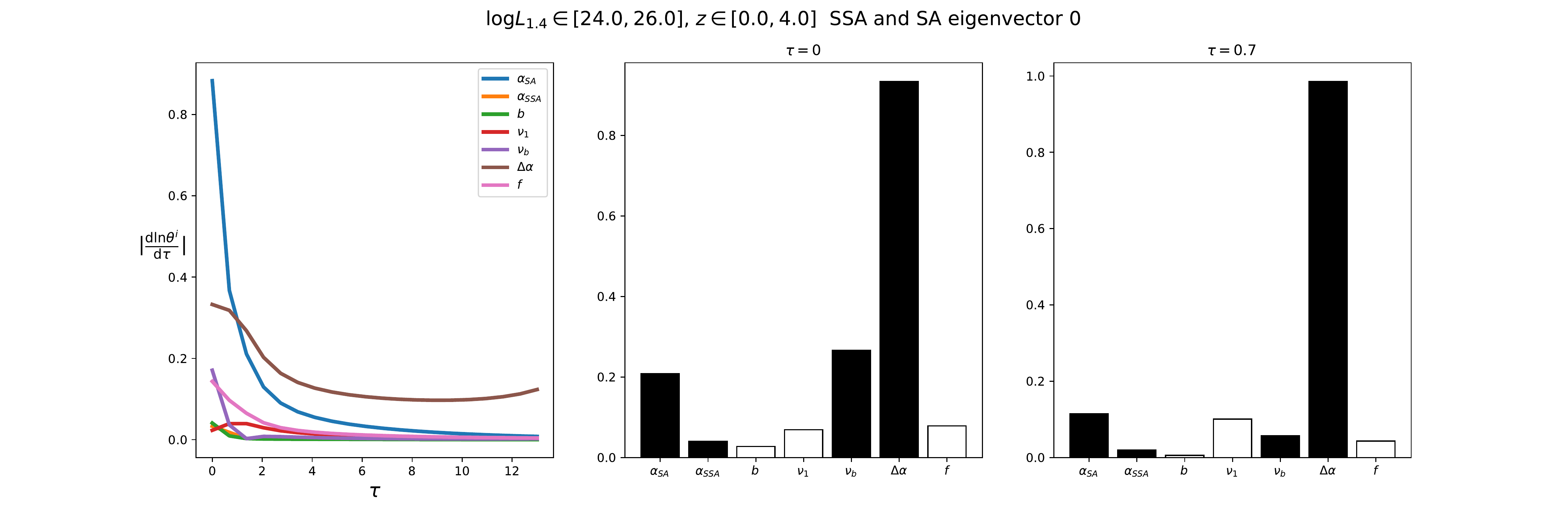}
  \caption{{Behavior of logarithmic derivative of parameters along the geodesic curve (left panel) and the components of the \acrshort{FIM} eigenvector corresponding to the smallest eigenvalue at the beginning (end) of the geodesic curve are shown in the middle (right) panel the model from Eq. \ref{eq:SSA+SA}. {The figure shows the details of the \acrshort{MBAM} method used in constraining the parameters of this complex model.} }}
  \label{fig:MBAM2}
\end{figure*}}
\newcommand{\FigLz}{
\begin{figure}[t]
  \centering
  \includegraphics[width=\linewidth]{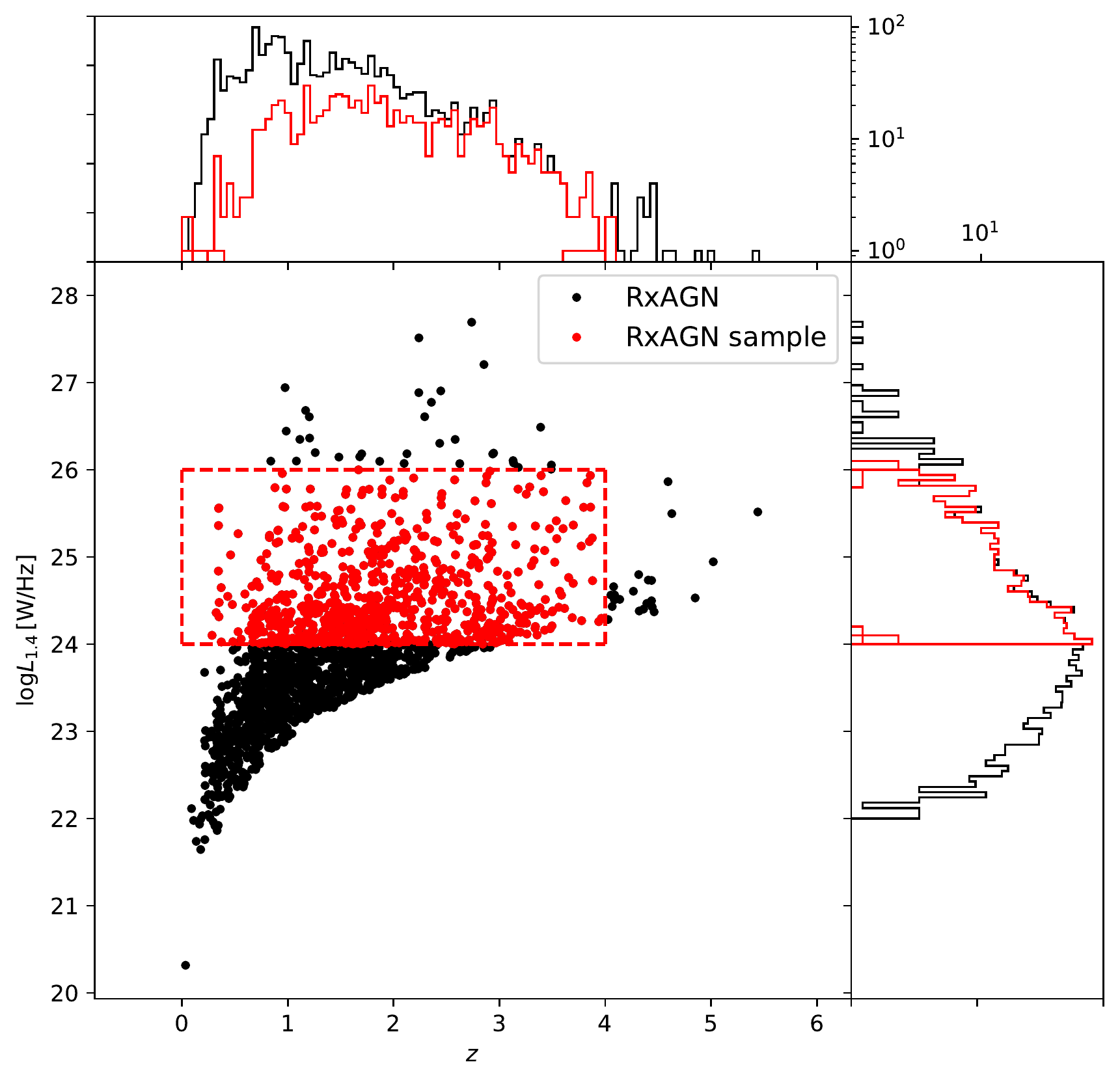}
  \caption{Radio luminosity {density} $L_{1.4}\,\mathrm{{[W/Hz]}}$ and redshift for \acrshort{RxAGN} (black points). Red points correspond to the selected redshift and radio luminosity sample of \acrshort{RxAGN} {used in this work and are contained within $\log L_{1.4}\,\mathrm{{[W/Hz]}}\in [24,26]$ and $z\in[0,4]$.} {The histograms show the distribution of radio luminosity and redshift for all \acrshort{RxAGN} and the \acrshort{RxAGN} sample {used in this work} with black and red bars, respectively.} {The radio luminosity-redshift distribution is used to choose a complete sample of \acrshort{RxAGN}.}}
  \label{fig:lgLz2d}
  \includegraphics[width=\linewidth]{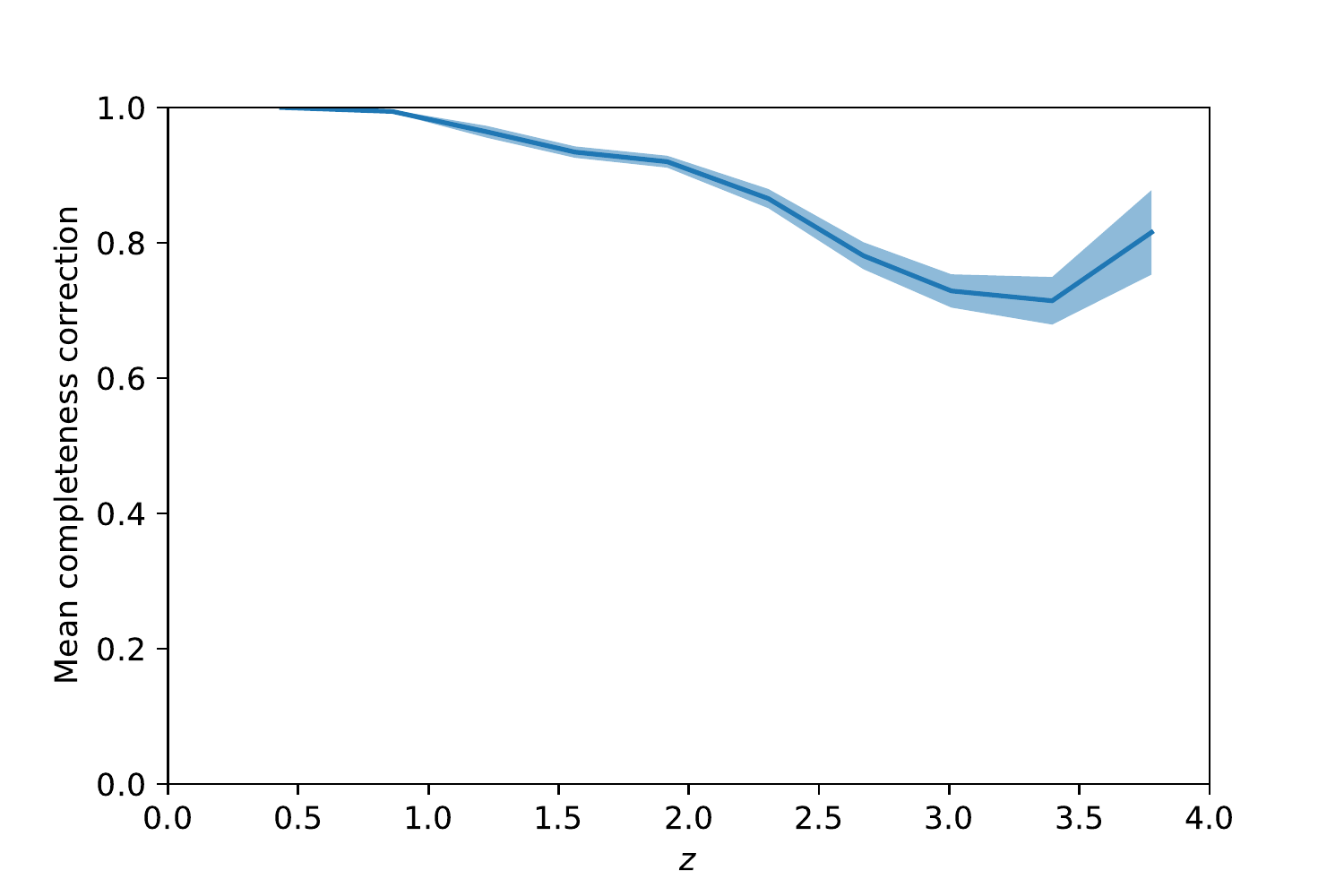}
  \caption{{Mean completeness correction for the sample of \acrshort{RxAGN} as a function of redshift. The mean completeness (solid line), with its corresponding standard deviation (shaded interval), were estimated for different redshift bins using the $3\GHz$ flux densities and interpolated completeness corrections for the \acrlong{VLA3LP} catalog \citep[table 2 in][]{Smolcic:17a}}. {The mean completeness correction  is used to quantify the completeness in the \gls{RxAGN} sample, which is higher than $75\%$ up to $z\sim 4$.}}
  \label{fig:Completeness_corr}
\end{figure}}
\newcommand{\FigSplitBins}{
\begin{figure*}
  \centering
  \includegraphics[width=2\columnwidth]{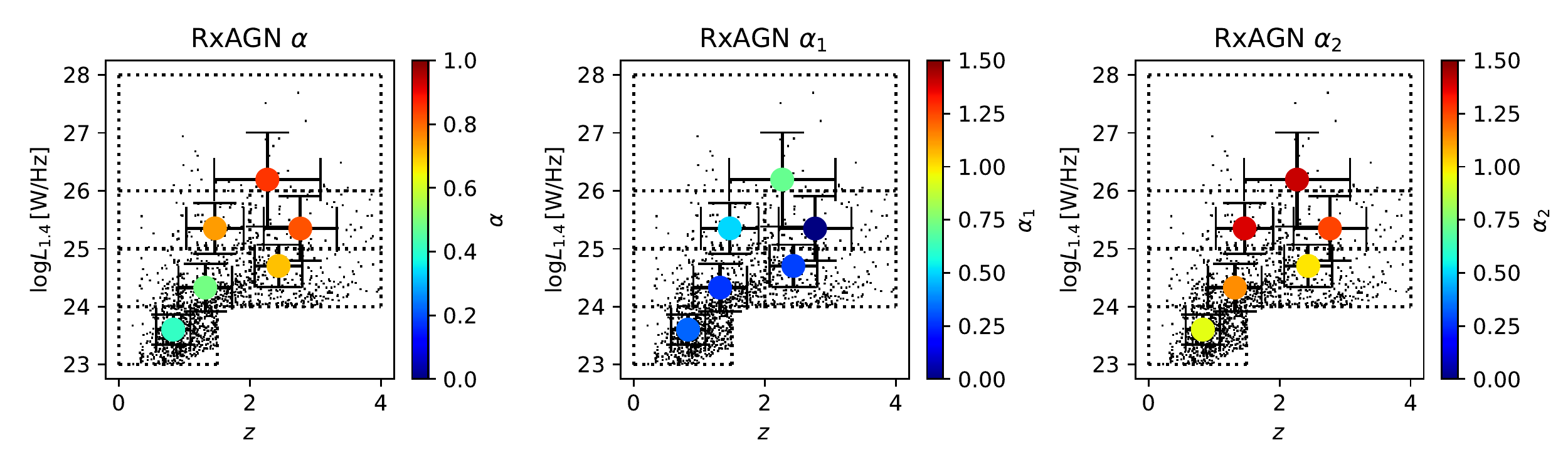}
  
  \caption{{Spectral indices derived for subsets of the \acrshort{RxAGN} sample divided by redshift, $z$, and radio luminosity, $\log L_{1.4}\,\mathrm{{[W/Hz]}}$, described in Sect. \ref{sect:DiscussionTrends}.} The color-scale in the left panel shows the spectral index of the \acrshort{PL} model, while the color-scales in the middle and right panels show the spectral indices $\alpha_1$ and $\alpha_2$ of the \acrlong{BPL} model. Bins are outlined by black dashed lines. {The panels show the spectral indices radio SEDs shown in Fig. \ref{fig:split_PL} for different subsets of \acrshort{RxAGN} and correspond to numbered bins presented in table \ref{tab:BPLresults} and Fig. \ref{fig:corrplot}.}}
  \label{fig:split_PL_bins}
\end{figure*}}
\newcommand{\FigCorrFirst}{\begin{figure*}
\begin{minipage}[t]{0.45\textwidth}
\includegraphics[width=\columnwidth]{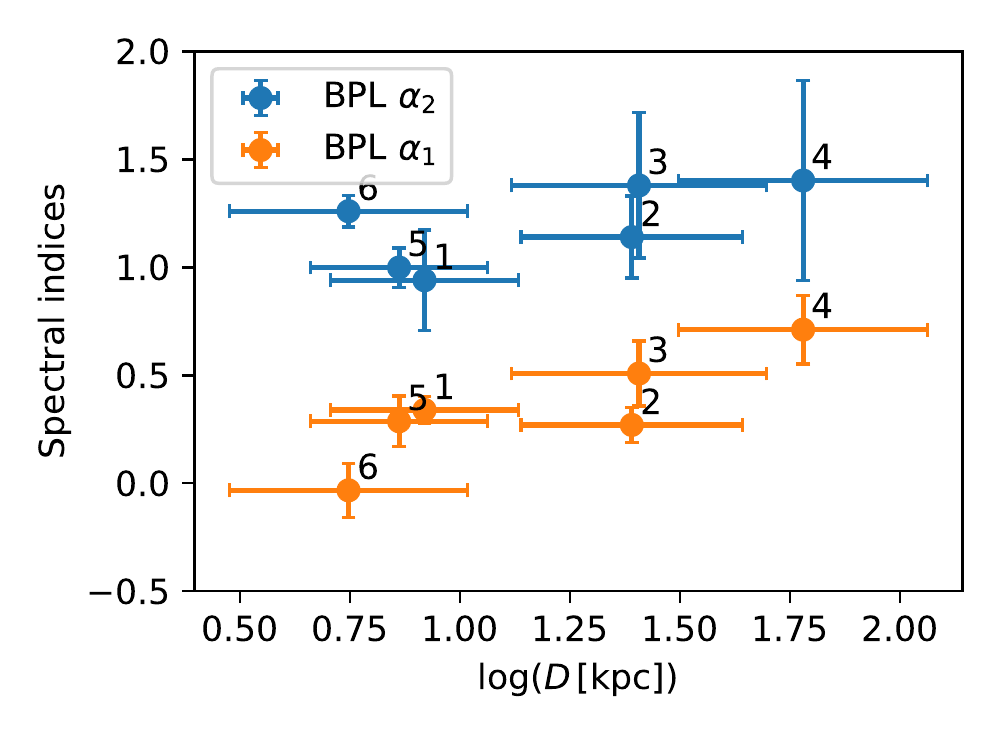}
\caption{{Dependence of the \acrlong{BPL} spectral indices below and above a break frequency of \BPLBreakFixedFrequency ($\alpha_1$ and $\alpha_2$, respectively) on the source size, $\log D\,\mathrm{{[kpc]}}$, }for the subsets of the \gls{RxAGN} {dataset for different values of in $\log D\,\mathrm{{[kpc]}}$, } with corresponding errors in spectral indices. The error bars for $\log D\,\mathrm{{[kpc]}}$ show the standard deviations of $\log D\,\mathrm{{[kpc]}}$ within each bin. The numbers indicate the bin number in table \ref{tab:BPLresults}. {The table lists the spectral indices radio SEDs shown in Fig. \ref{fig:split_PL} for different subsets of \acrshort{RxAGN} and correspond to numbered bins presented in Fig. \ref{fig:corrplot}.}}\label{fig:corrplot}
\end{minipage}\hfill
\begin{minipage}[t]{\columnwidth}
\includegraphics[width=\columnwidth]{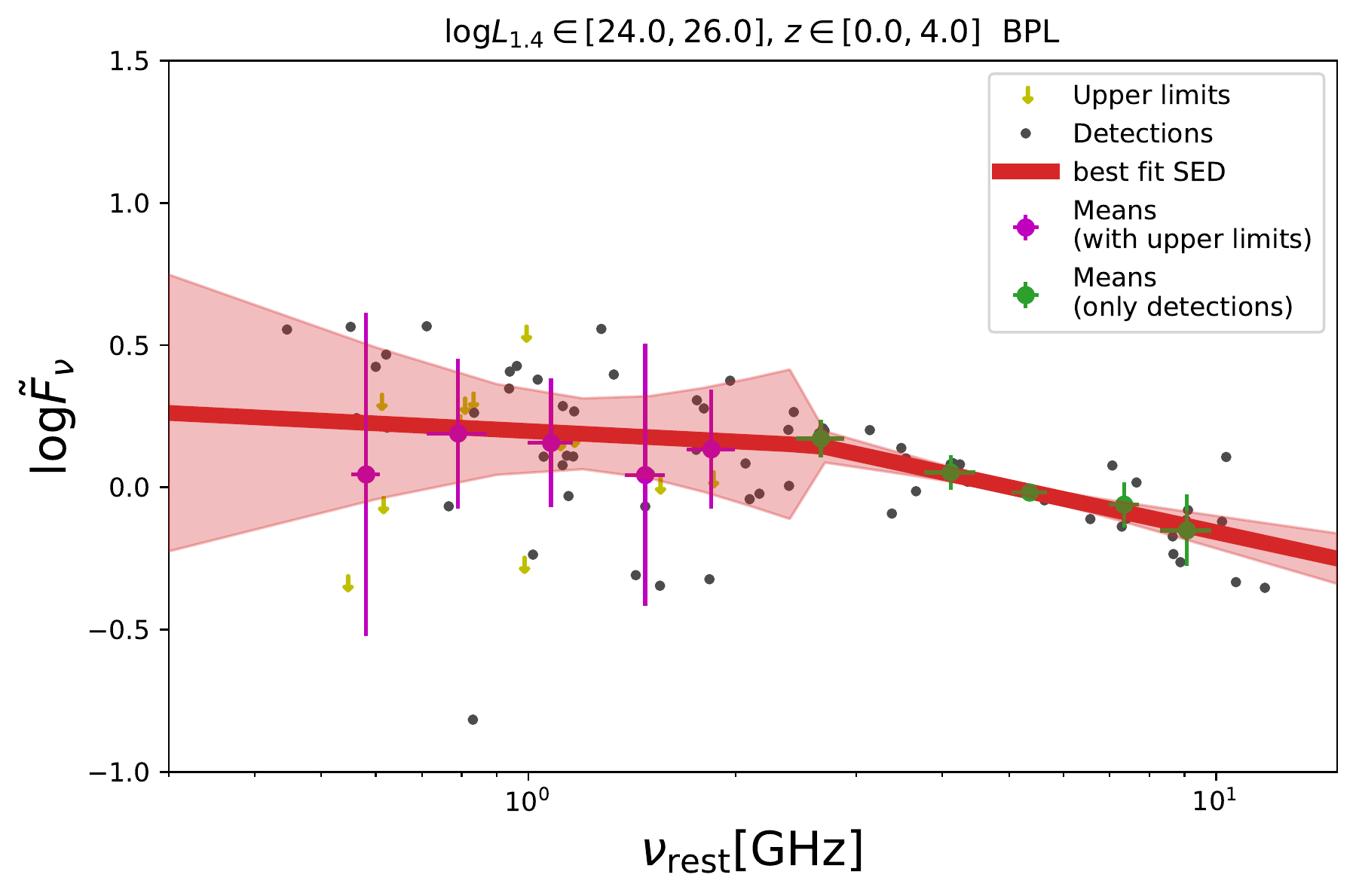}
\caption{{Radio} \acrshort{SED} of a sample of flat-spectrum sources, produced by {selecting all sources having sizes less than $1\kpc$}. Red line shows the \acrlong{BPL} model. {We find that selecting sources with sizes less than $1\kpc$ produces a sample of flat-spectrum sources, described by a spectral index of \flatPL and a broken power-law spectral index of \flatBPLlow (\flatBPLhigh) below (above) a break frequency of \flatBPLFrequency.}}\label{fig:inv}
\end{minipage}

\begin{minipage}[t]{\columnwidth}
\centering
\includegraphics[width=.95\columnwidth]{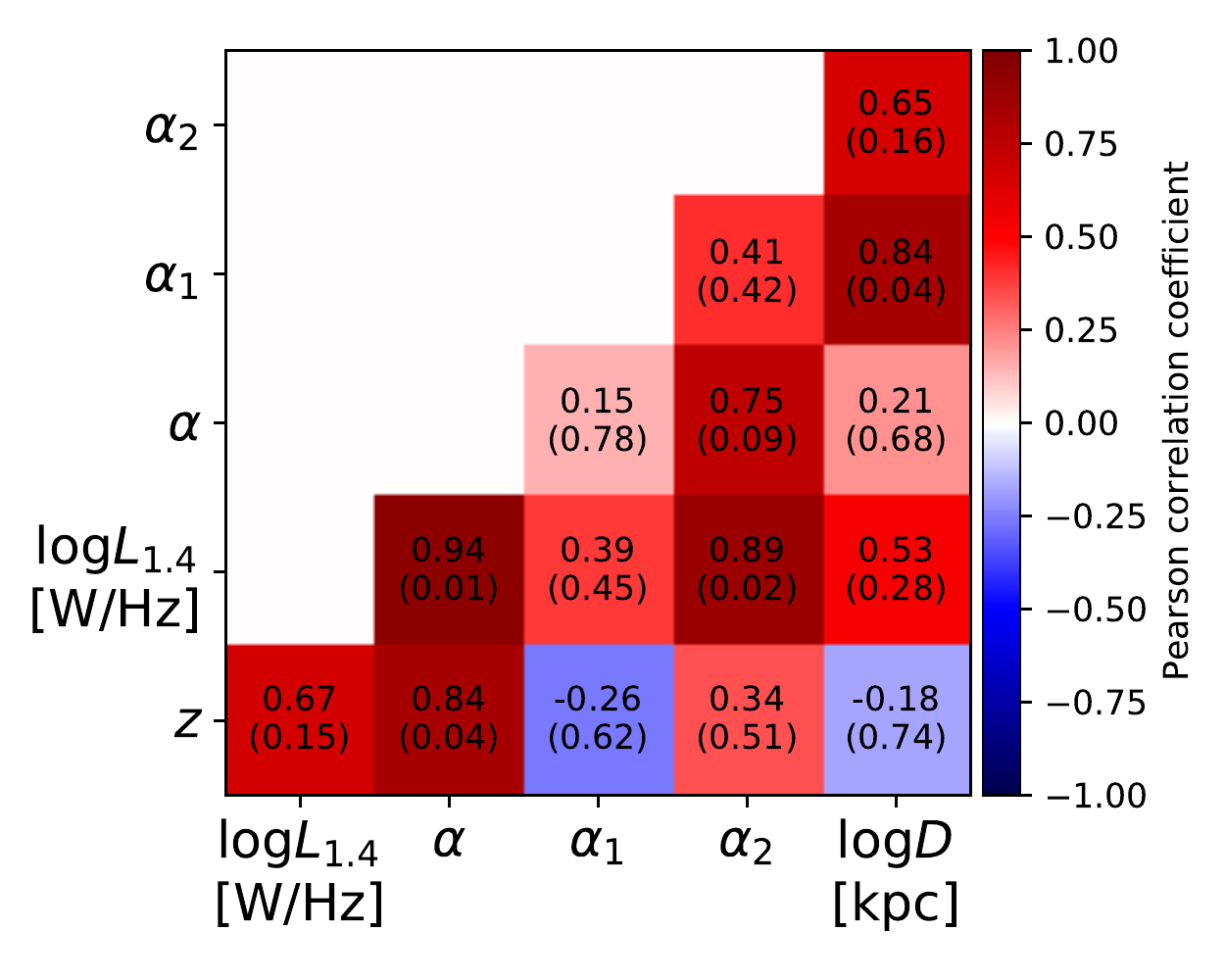}
\caption{Correlation matrix for properties of subsets of the \acrshort{RxAGN} sample. Colors indicate positive or negative Pearson correlation coefficient, written on the plot for each parameter pair with its corresponding p-value in parentheses. {The considered variables in the correlation matrix are the $1.4\GHz$ radio luminosity, $\log L_{1.4}\,\mathrm{{[W/Hz]}}$, \acrlong{PL} spectral index $\alpha$, \acrlong{BPL} spectral indices $\alpha_1$ and $\alpha_2$, and source size, $\log D\,\mathrm{{[kpc]}}$.} {The correlation matrix presented in this figure indicates the presence and significance of correlations between different properties of the subsets of the RxAGN sample.}}\label{fig:corrplot2}
\end{minipage}\hfill
\begin{minipage}[t]{\columnwidth}
\includegraphics[width=\columnwidth]{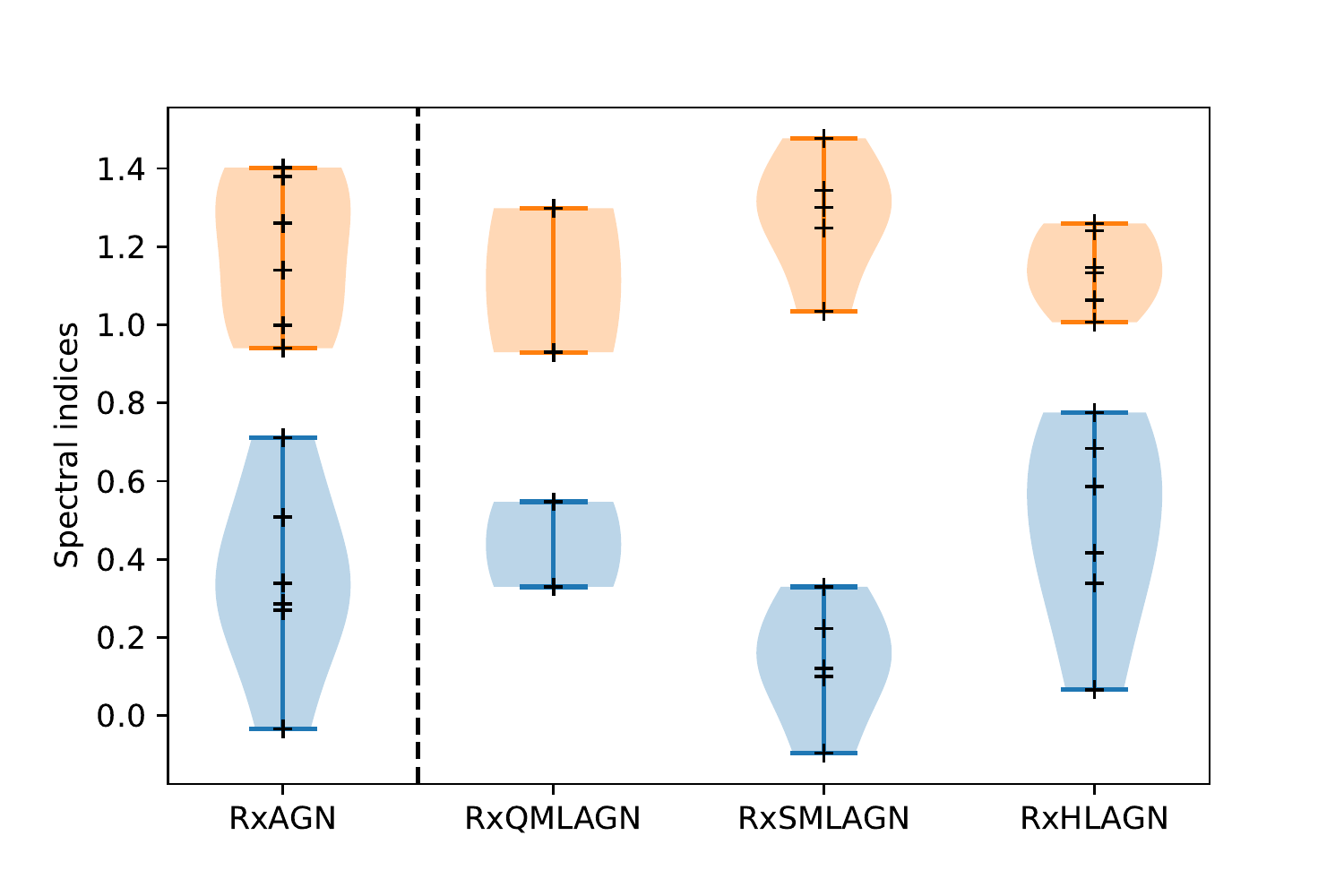}
\caption{Distribution of the \acrlong{BPL} spectral indices for the subsets of the \acrshort{RxAGN} sample. {Black crosses are spectral indices derived for individual bins, listed in Tab. \ref{tab:BPLresults}}. Blue intervals denote $\alpha_1$ while orange intervals denote $\alpha_2$. {The distributions of the \acrlong{BPL} spectral indices are used to determine the parameters that explain the differences between the \acrshort{RxAGN} subsets using the ANOVA method summarized in table \ref{tab:ANOVA}. }}\label{fig:violinplot}
\end{minipage}
\end{figure*}}
\newcommand{\FigSplitSEDs}{\begin{figure*}
\includegraphics[width=\columnwidth]{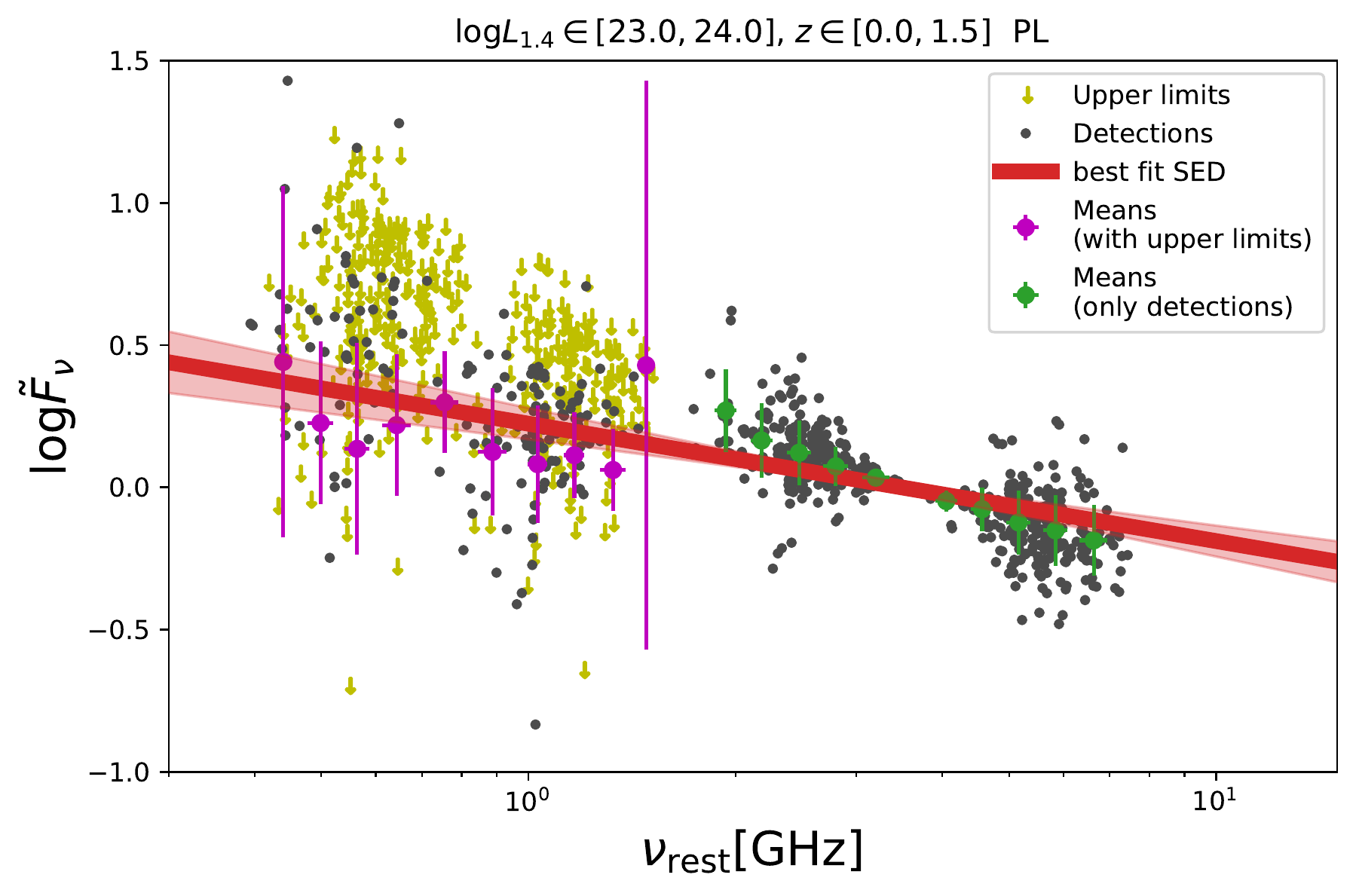}
\includegraphics[width=\columnwidth]{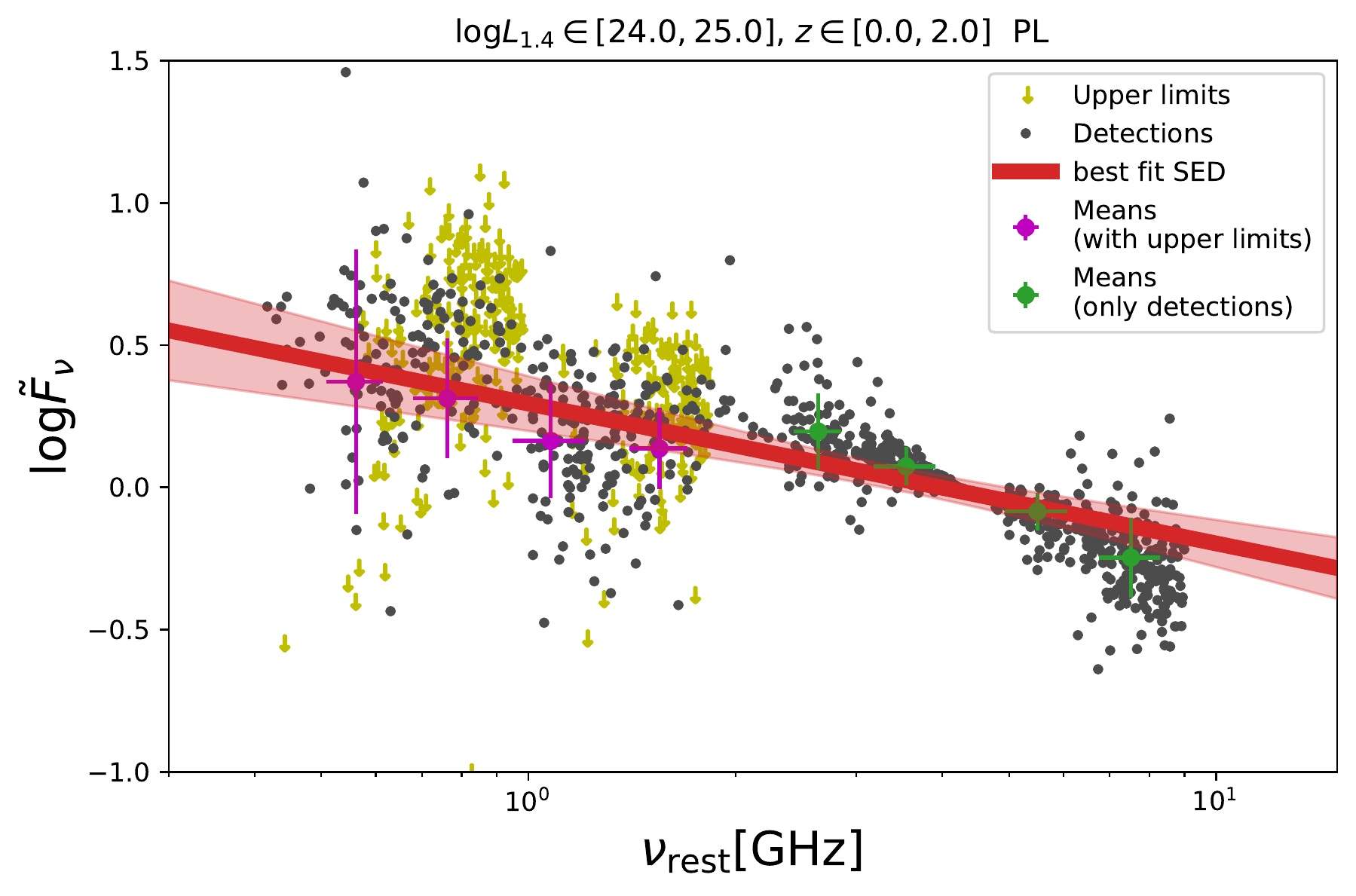}

\includegraphics[width=\columnwidth]{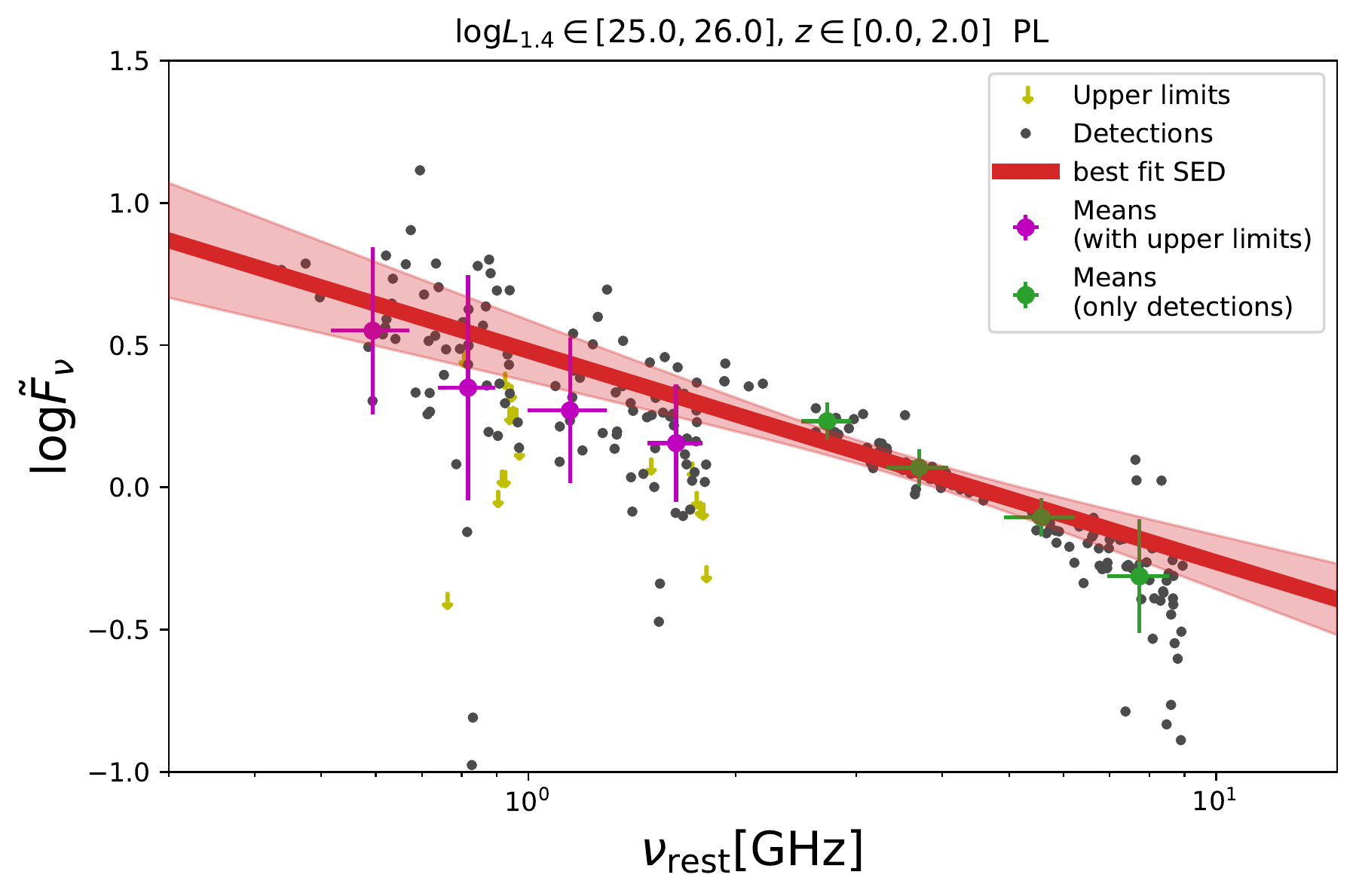}
\includegraphics[width=\columnwidth]{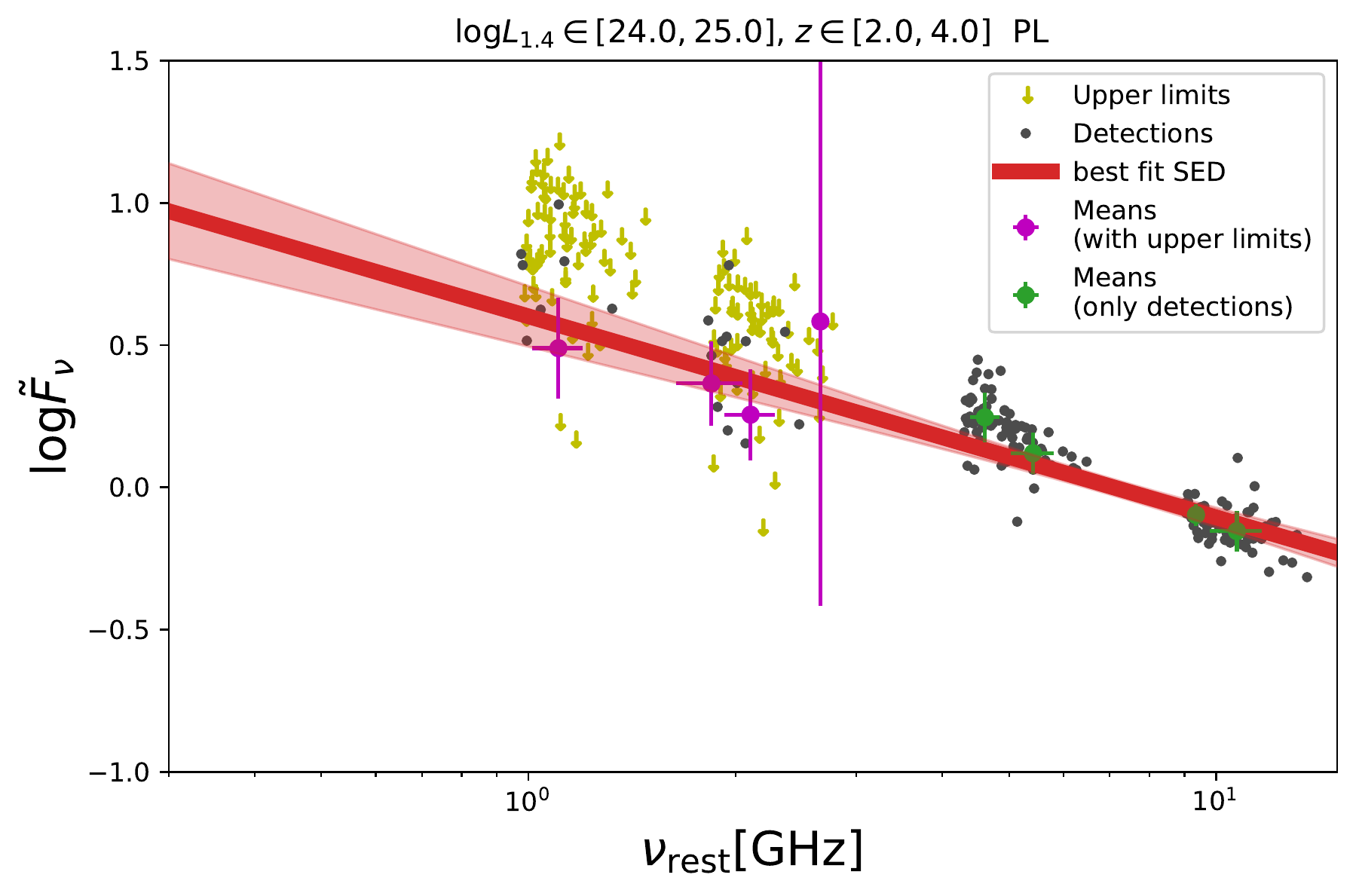}

\includegraphics[width=\columnwidth]{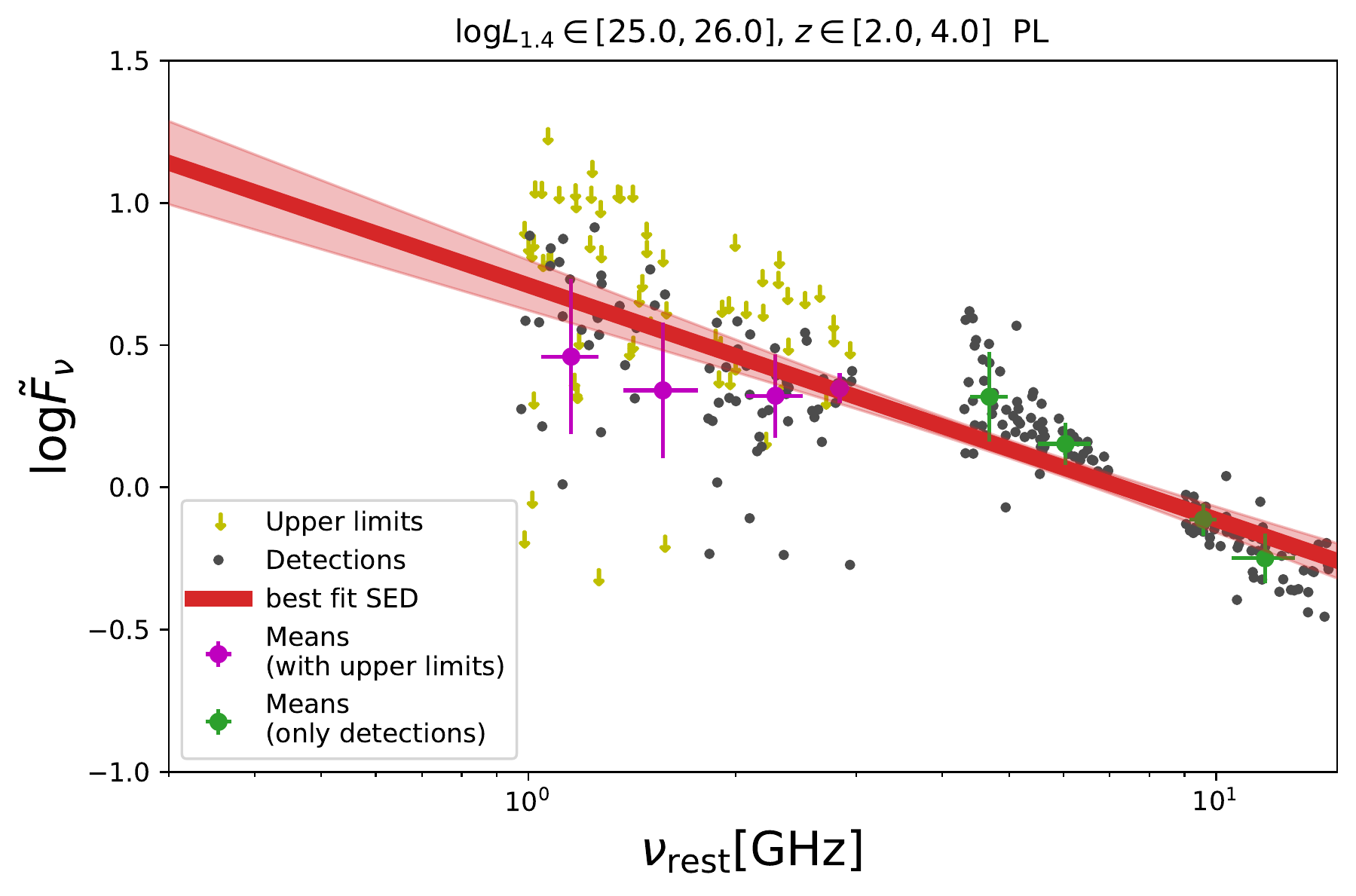}
\includegraphics[width=\columnwidth]{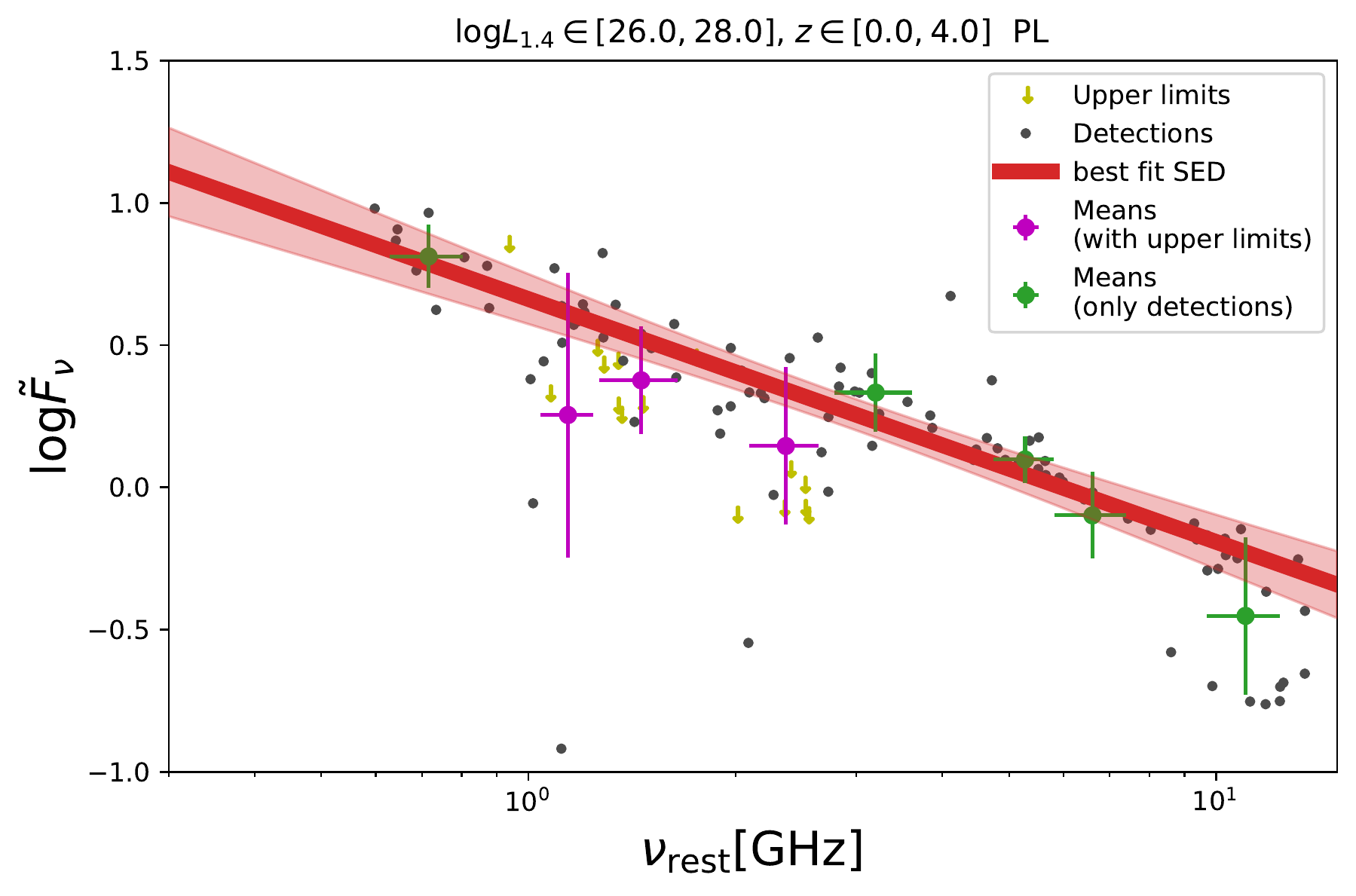}
\caption{\acrshort{SED}s derived for subsets of the \acrshort{RxAGN} sample divided by redshift, $z$, and radio luminosity, $\log L_{1.4}\,\mathrm{{[W/Hz]}}$. Shown is the best-fitting power-law model for each bin. {The panels show the radio SEDs for different subsets of \acrshort{RxAGN} and correspond to numbered bins presented in table \ref{tab:BPLresults} and Fig. \ref{fig:corrplot}.}}\label{fig:split_PL}
\end{figure*}}
\newcommand{\TabANOVA}{\begin{table}[]
  \centering
  \caption{Summary of ANOVA results for the \acrlong{PL} and \acrlong{BPL} spectral indices for \acrshort{RxAGN} subsamples for models selected  by requiring $P<0.1$. { For each model the significance of dependence of the spectral index on each of the listed parameters is determined from the P-value. The considered parameters are: Type- if the sources in the bins were \gls{RxHLAGN}, \gls{RxSMLAGN} or \gls{RxQMLAGN}, $z$ - redshift, and $\log D\,\mathrm{{[kpc]}}$ - source size.}}
  \begin{tabular}{c c c c}
  \toprule
  \toprule
  Model & Parameter &  P-value\\
  \midrule
  $\alpha(z)$ & $z$ & $8\times 10^{-5}$\\
  \midrule
  $\alpha_1(z)$ & $z$ & 0.092\\
  \midrule
  $\alpha_1(\mathrm{Type},  \log D\,\mathrm{{[kpc]}})$ & RxSMLAGN & 0.078\\
            &$\log D\,\mathrm{{[kpc]}}$ & 0.017\\
\midrule
$\alpha_2(z, \log D\,\mathrm{{[kpc]}})$ &$z$ &  0.006\\
& $\log D\,\mathrm{{[kpc]}}$ & 0.008\\
 \bottomrule
  \end{tabular}

  \label{tab:ANOVA}
\end{table}}
\newcommand{\TabIndices}{\begin{table*}[]
  \centering
  \caption{Parameters of the \acrlong{PL} and \acrlong{BPL} models, {and source size, $\log D\,\mathrm{{[kpc]}}$}, derived for subsets of the \acrshort{RxAGN} sample divided by redshift, $z$, and radio luminosity, $\log L_{1.4}\,\mathrm{{[W/Hz]}}$, {described by the mean and standard deviation of redshifts and radio luminosity for each bin} The sample is further split into \acrshort{RxHLAGN}, \acrshort{RxQMLAGN}, and \acrshort{RxSMLAGN}. {The numbers 1-6 in parentheses indicate bin labels in Fig. \ref{fig:corrplot}.}}
  \label{tab:BPLresults}
  \begin{tabular}{c c c c c c c}
  \toprule
  \toprule
   $z$ &$\log L_{1.4}\,\mathrm{{[W/Hz]}}$ &Type & $\alpha$ &$\alpha_1$ & $\alpha_2$ & $\log D\,\mathrm{{[kpc]}}$\\
\midrule
 $0.83\pm0.26$ & $23.60\pm0.28$ & RxAGN (1) & $0.41\pm0.06$ & $0.34\pm0.06$ & $0.94\pm0.23$ & $0.92\pm0.21$\\
 $1.3\pm0.4$ & $24.33\pm0.24$ & RxAGN (2) & $0.49\pm0.10$ & $0.27\pm0.08$ & $1.14\pm0.19$ & $1.39\pm0.25$\\
 $1.5\pm0.4$ & $25.35\pm0.26$ & RxAGN (3) & $0.74\pm0.13$ & $0.51\pm0.15$ & $1.38\pm0.34$ & $1.41\pm0.29$\\
 $2.3\pm0.8$ & $26.2\pm0.4$ & RxAGN (4) & $0.85\pm0.11$ & $0.71\pm0.16$ & $1.4\pm0.5$ & $1.78\pm0.28$\\
 $2.4\pm0.4$ & $24.70\pm0.21$ & RxAGN (5)& $0.71\pm0.09$ & $0.29\pm0.12$ & $1.00\pm0.09$ & $0.86\pm0.20$\\
 $2.8\pm0.6$ & $25.35\pm0.28$ & RxAGN (6) & $0.82\pm0.10$ & $-0.03\pm0.13$ & $1.26\pm0.07$ & $0.75\pm0.27$\\
 $1.0\pm0.4$ & $24.29\pm0.25$ & RxQMLAGN & $0.50\pm0.09$ & $0.33\pm0.10$ & $0.93\pm0.22$ & $1.12\pm0.27$\\
 $1.2\pm0.4$ & $25.28\pm0.22$ & RxQMLAGN & $0.75\pm0.11$ & $0.55\pm0.16$ & $1.3\pm0.4$ & $1.13\pm0.31$\\
 $1.5\pm0.4$ & $24.34\pm0.24$ & RxSMLAGN & $0.56\pm0.11$ & $0.33\pm0.11$ & $1.25\pm0.27$ & $1.40\pm0.24$\\
 $1.77\pm0.30$ & $25.27\pm0.29$ & RxSMLAGN & $0.79\pm0.10$ & $0.22\pm0.10$ & $1.34\pm0.10$ & $1.54\pm0.30$\\
 $2.6\pm0.6$ & $26.1\pm0.4$ & RxSMLAGN & $0.82\pm0.17$ & $0.1\pm0.4$ & $1.48\pm0.35$ & $1.19\pm0.30$\\
 $2.46\pm0.34$ & $24.71\pm0.21$ & RxSMLAGN & $0.72\pm0.12$ & $0.10\pm0.12$ & $1.03\pm0.07$ & $0.86\pm0.20$\\
 $2.7\pm0.6$ & $25.32\pm0.26$ & RxSMLAGN & $0.79\pm0.12$ & $-0.10\pm0.15$ & $1.30\pm0.10$ & $1.06\pm0.27$\\
 $0.82\pm0.25$ & $23.69\pm0.27$ & RxHLAGN & $0.50\pm0.07$ & $0.42\pm0.08$ & $1.01\pm0.28$ & $1.11\pm0.23$\\
 $1.3\pm0.4$ & $24.35\pm0.24$ & RxHLAGN & $0.55\pm0.10$ & $0.34\pm0.06$ & $1.15\pm0.14$ & $1.89\pm0.27$\\
 $1.4\pm0.5$ & $25.37\pm0.25$ & RxHLAGN & $0.78\pm0.07$ & $0.68\pm0.08$ & $1.13\pm0.23$ & $1.30\pm0.30$\\
 $2.2\pm0.8$ & $26.2\pm0.5$ & RxHLAGN & $0.89\pm0.08$ & $0.78\pm0.11$ & $1.24\pm0.27$ & $1.54\pm0.29$\\
 $2.4\pm0.4$ & $24.68\pm0.16$ & RxHLAGN & $0.89\pm0.07$ & $0.59\pm0.10$ & $1.06\pm0.07$ & $0.78\pm0.21$\\
 $2.9\pm0.6$ & $25.39\pm0.30$ & RxHLAGN & $0.85\pm0.09$ & $0.07\pm0.13$ & $1.26\pm0.07$ & $0.60\pm0.28$\\
 \bottomrule
  \end{tabular}

\end{table*}}
\usepackage{etoolbox}
\makeatletter
\patchcmd\@combinedblfloats{\box\@outputbox}{\unvbox\@outputbox}{}{\errmessage{\noexpand patch failed}}
\makeatother
\begin{document}

\abstract { As the Square Kilometer Array (SKA) is expected to be operational in the next decade, investigations of the radio sky in the range of 100 MHz to 10 GHz have become important for {simulating SKA observations}. In determining physical properties of galaxies from radio data, the radio spectral energy distribution (SED) is often assumed to be described by a simple {power law}, usually with a spectral index of 0.7 for all sources. Even though radio SEDs have been shown to exhibit deviations from this assumption, both in differing spectral indices and complex spectral shapes, it is often presumed that their individual differences {can be canceled} out in large samples. } {{Since the average spectral index around $1\GHz$ (observed-frame) is important for determining physical properties of large samples of galaxies, we aim to test whether individual differences in the spectra of radio-identified active galactic nuclei align with the simple assumption of $\alpha=0.7$ and test the evolution of the parameters of the synchrotron aging model with redshift and radio luminosity.} } {{We use a sample of 744 radio-excess active galactic nuclei (RxAGN), defined as those that exhibit more than a $3\sigma$ radio luminosity excess with respect to the value expected only from {the} contribution from star formation, out to $z\sim 4$. We {constructed} their average radio SED by combining Very Large Array (VLA) observations of the COSMOS field at $1.4\GHz$ and $3\GHz$ with Giant Meterwave Radio Telescope (GMRT) observations at $325\MHz$ and $610\MHz$.} To account for {nondetections} in the GMRT maps, we employed the survival analysis technique. We binned the RxAGN sample into luminosity- and redshift-complete subsamples. In each bin, we constrained the shape of the average radio SED by fitting a broken power-law model. } {We find that the RxAGN sample can be described by a spectral index of \BPLSpectralIndexLower below the break frequency \BPLBreakFrequency and \BPLSpectralIndexHigher above it, {while a simple power-law model, {capturing fewer spectral features}, yields a single spectral index of \PLSpectralIndex. {By binning in $1.4\GHz$ {of} radio luminosity and redshift, we find that the power-law spectral index is positively correlated with redshift and that the broken power-law spectral index above \BPLBreakFixedFrequency is positively correlated with both {the} redshift and source size.} By selecting sources with sizes less than $1\kpc$, we find a subsample of flat-spectrum sources, {which can be} described by a spectral index of \flatPL and a broken power-law spectral index of \flatBPLlow (\flatBPLhigh) below (above) a break frequency of \flatBPLFrequency.}} {{ We have constrained the radio SED for a sample of RxAGN in the COSMOS field using available VLA and GMRT data, {corresponding  to} the rest-frame frequency range from $\sim 0.3\GHz$ to $\sim 10\GHz$. We describe our derived average radio SED of RxAGN using power-law and broken power-law models, yielding a radio SED that steepens above $\sim 4\GHz$. }}
\maketitle
\section{Introduction}
In recent years, {the} investigation of the radio sky at sub-GHz frequencies has received more attention \citep[see e.g., ][]{Padovani2016TheMainstream} since the advent of \gls{LOFAR}, which can operate in the $15-200\MHz$ frequency range \citep{vanHaarlem13}. Moreover, the \gls{SKA} is expected to operate at frequencies from $50\MHz$ to $20\GHz$\footnote{\url{https://www.skatelescope.org/wp-content/uploads/2018/08/16231-factsheet-telescopes-v71.pdf}} with nanojansky sensitivities \citep{Norris2013RadioPathfinders}, {while precursors, {such as} \gls{MWACS}, have already started} producing catalogs of sources detected at $\sim100\MHz$. Our ability to make predictions {for} these new instruments is tied to the knowledge of the \gls{SED} of sources upon which semi-empirical simulations are devised \citep{Wilman2008ATelescopes}.

The radio \gls{SED} of sources in the frequency range of $1-10\,\mathrm{GHz}$ are often assumed to be described by a simple {power law}, with a spectral index of $\alpha=0.7$ \citep[{using the convention $S\sim\nu^{-\alpha}$,}][]{Condon92}. The simple power-law shape of the radio \gls{SED} is expected to be mainly due to synchrotron radiation, {arising either from supernova remnants, tracing star formation, or from the vicinity of supermassive black holes, tracing \gls{AGN} activity.}
However, the radio \gls{SED}s have been shown to exhibit deviations from these simple relationships by having different spectral indices and complex spectral shapes \citep[see, e.g.,][]{Condon1991b,Kukula98, Kimball08,Clemens08, Clemens2010,Leroy11,Murphy2013TheGalaxies, CalistroRivera17, Galvin17, Tisanic19}. For example, quasars have been shown to have spectral indices in the $1-10\GHz$ range {which} may be either steep (>0.5) or flat (<0.5) \citep{Kukula98}.
At $150\MHz$, \citet{Toba19} find an even greater variety in radio spectral indices, ranging from $0.5$ to $2.5$. 

{In the simple picture,} the shape of the radio \gls{SED} of individual sources at higher frequencies is related to the synchrotron aging process and additionally to free-free emission in star-forming galaxies, while the lower-frequency part of the spectrum may be influenced by the different absorption processes \citep[see, e.g., ][]{Condon92}. {In bright samples ($\sim 1 \Jy$), alongside flat-spectrum sources $(\alpha<0.5$), a significant fraction ($~40-50\%$) shows either a steeper \gls{SED} {($\alpha>0.5$), or a peaked or inverted spectrum, \citep[][]{Kapahi1981WesterborkSurvey.,Peacock1982BrightTelescope,deZotti10}}.} \citet{ Kimball08} found that individual \gls{SED}s of complex and resolved sources (at a $5\arcsec$ resolution) are best described by a spectral index of $\sim 0.8$, while unresolved sources have a flat spectral shape ($\alpha\sim0$). They point out that emission from extended radio lobes tends to be steeper, while flatter emission might arise from compact quasar cores {and} jets due to self-absorbed synchrotron emission. Synchrotron aging is usually described either assuming a single injection of electrons with a constant or isotropic pitch angle, or assuming continuous injection of electrons \citep{Kardashev62, JP73,Pacholczyk80}. Absorption processes affecting the \gls{SED} shape are synchrotron self-absorption and free-free absorption, which have been used to explain radio spectra of gigahertz peaked-spectrum and compact steep-spectrum sources \citep{ODea98, Collier2018High-resolutionSources}. \citet{ODea97} found that the source size for these types of sources is negatively correlated with turnover frequency in their spectra \citep{Fanti1990OnSources,ODea97}, while sources have been found with turnover frequencies above $10\GHz$ \citep{Edge1998GPSFrequencies}. 
 By using \gls{LOFAR} observations, \citet{CalistroRivera17} find statistically significant steepening in their \gls{AGN} \gls{SED} in the $150\MHz-600\MHz$ range compared to \gls{AGN} \gls{SED}s around $1\GHz$. 

{Previous studies of radio \gls{SED}s of individual sources find that the average value of the spectral index is consistent with the expected value of $\alpha=0.7$} \citep[e.g.,][]{Kukula98, Toba19}.
 Radio luminosity functions of \gls{AGN} have been shown to differ for flat and steep spectrum sources \citep{Dunlop1990TheQuasars.}. \citet{Novak18} {quantify the impact of different \gls{SED} shapes on radio luminosity functions by pointing out} that while the differences in individual galaxies' \gls{SED}s {can be canceled} out for large samples, even a slight change of $0.1$ in the mean value of the spectral index could result in a $0.1\dex$ change in rest-frame luminosities.

 In this paper we present a survival analysis study of the average radio \gls{SED} of a sample of {radio-excess} \gls{AGN} in the \gls{COSMOS} field, identified by an excess in radio luminosity compared to that expected {solely from star-forming} processes. {A combination of radio data available in the \gls{COSMOS} field offers a way to construct approximately luminosity-complete samples out to $z\sim 4$, farther than would be possible when considering sources individually. We construct average radio \gls{SED}s following the method detailed in \citet{Tisanic19}.}
 
In Sect. \ref{sect:Data} we describe the available observations and data used in the constriction of the radio \gls{SED}s and the selection of the radio-excess \gls{AGN} sample. In Sect. \ref{sect:Methods} we briefly summarize the method used to construct the radio \gls{SED}s and describe models used to analyze its shape. We use the convention that a steeper \gls{SED} means a larger spectral index ($S\sim\nu^{-\alpha}$) throughout this paper and adopt the following cosmological parameters: $\Omega_M=0.3$, $\Omega_\Lambda=0.7$, and $H_0=70\,\mathrm{km/s/Mpc}$.
\section{Data and sample}\label{sect:Data}

To assess the radio \gls{SED}, we combined catalogs of the {radio data in the} \gls{COSMOS} field at four observer-frame frequencies: $325\MHz$ and $610\MHz$, obtained with the \gls{GMRT} \citep{Tisanic19}, and at $1.4\GHz$ 
\citep{Schinnerer10} $\text{and }3\GHz$ \citep{Smolcic:17a}, obtained {with} the \gls{VLA}.
Here we briefly describe the \gls{VLA} (Sect. \ref{sect:DataVLA}) and \gls{GMRT} (Sect. \ref{sect:DataGMRT}) data and {the selection of} the sample used in this paper (Sect. \ref{sect:DataSample}).

\subsection{VLA data}\label{sect:DataVLA}
We have combined catalogs of the \gls{VLA3LP} and the \gls{VLA1.4JP} .

The \gls{VLA3LP} map was constructed from $384\hr$ of observations of the $2\deg^2$ \gls{COSMOS} field. 
Observations were carried out over $192$ pointings in the S-band with \gls{VLA} in A and C antenna configurations.
These were wide-band observations with a total bandwidth of $2084\MHz$ derived from $16$ spectral windows, each $128\MHz$ wide. The final mosaic reached a median \gls{RMS} of $2.3\muJy/\mathrm{beam}$ at a resolution of $0.75\arcsec$.
Considering the large bandwidth and the volume of the data, each pointing was imaged separately using the \gls{MSMF} and then combined into a single mosaic. 
The \gls{MSMF} algorithm had been found to be optimal for both resolution and image quality in terms of the \gls{RMS} noise and sidelobe contamination \citep{Novak15}.
The \gls{VLA3LP} catalog has been derived using \textsc{blobcat} \citep{Hales12} with a $5\sigma$ threshold.
In total, 10830 sources were recovered (67 of which were multi-component sources). For more details, see \citet{Smolcic:17a}.

The \gls{VLA1.4JP} catalog is a joint catalog comprised of the \gls{VLA1.4LP} and \gls{VLA1.4DP} surveys, as described in \citet{Schinnerer10}. The \gls{VLA1.4JP} catalog was constructed by combining the observations of the entire $2\deg^2$ \gls{COSMOS} field at a resolution of $1.5\arcsec\times1.4\arcsec$ \citep[][]{Schinnerer07}, and observations of the central $50\arcmin\times 50\arcmin$ subregion of the \gls{COSMOS} field at a resolution of $2.5\arcsec\times2.5\arcsec$ \citep[][]{Schinnerer10}. The average \gls{RMS} in the resulting map was found to be $12\muJy$.
The \gls{VLA1.4LP} observations consisted of $240\hr$ of observations over $23$ pointings spread out over the entire \gls{COSMOS} field in the L band centered at $1.4\GHz$ and with a total bandwidth of $37.5\MHz$ in \gls{VLA} A configuration and $24\hr$ in the C configuration. 
The \gls{VLA1.4DP} observations supplemented the $7$ central pointings with an additional $8.25\hr$ of observations per pointing using the A configuration and the same L-band configuration. 
These new measurements were then combined in the uv-plane with the \gls{VLA1.4LP} observations.
The \gls{VLA1.4JP} catalog was constructed by using the SExtractor package \citep{BertinArnouts96} and the AIPS task SAD, yielding 2865 sources \citep{Schinnerer10}.

\subsection{GMRT data}\label{sect:DataGMRT}

The $325\MHz$ observations of a single pointing were carried out {with the \gls{GMRT}} using a bandwidth of $32\MHz$ \citep{Tisanic19}.
The observations lasted for $45\hr$ in total, {comprising four observations with a total on-source time of $\sim40\hr$.} They were reduced using the \gls{SPAM} pipeline and imaged at a resolution of $10.8\arcsec\times 9.5\arcsec$. A primary beam correction was applied to the pointing.
{We {measured} a median \gls{RMS} of $97\muJy/\mathrm{beam}$ over the $\sim 2 \deg^2$ \gls{COSMOS} field.}
In total, 633 sources were identified using \textsc{blobcat} down to $5\sigma$. By employing a total-over-peak {flux density} criterion, we consider 177 of these sources to be resolved \citep[{derived by mirroring the fifth percentile of the total-over-peak flux density derived in bins of the signal-to-noise ratio, see} ][for details]{Tisanic19}.

The $610\MHz$ observations are described in detail in \citet{Tisanic19}. They were carried out at a central frequency of $608\MHz$ using a bandwidth of $32\MHz$.
Observations lasting $86\hr$ were conducted (spread over eight observations with an average time on source per pointing of $\sim 4\hr$) during which a total of $19$ pointings were observed. The data were reduced using the \gls{SPAM} pipeline and imaged at a resolution of $5.6\arcsec\times 3.9\arcsec$. A primary beam correction and average pointing error corrections were applied to each pointing prior to mosaicing.
{We {measured} a median \gls{RMS} of $39\muJy/\mathrm{beam}$ in the final mosaic over the $\sim 2 \deg^2$ \gls{COSMOS} field. }
\textsc{Blobcat} was used to extract 999 sources down to $5\sigma$, 196 of which we consider to be resolved {by a total-over-peak flux density criterion \citep[see ][and preceding paragraph for details]{Tisanic19}.}

\subsection{The 1.4 GHz selected sample}\label{sect:DataSample}
In order to obtain at least two data points {per source} for the \gls{SED} analysis, following \citet{Tisanic19} we have defined and analyzed a $1.4\GHz$-selected sample of sources from the \gls{VLA1.4JP} catalog that were detected in the \gls{VLA3LP} counterpart catalog. The $3\GHz$ data have lower \gls{RMS} ($\sim 2\,\mathrm{\mu Jy/beam}$) than the $\sim 12\,\mathrm{\mu Jy/beam}$ \gls{RMS} of the $1.4\GHz$ data. This procedure introduced only minimal potential biases stemming from the $1.4$ to $3\GHz$ spectral index distribution, as almost all $1.4\GHz$ sources \citep[$\sim 90\percent$,][]{Smolcic:17a} have counterparts at $3\GHz$. 

\citet{Smolcic:17b} obtained photometric redshifts by cross-correlating the $3\GHz$ Source Catalog with multi-wavelength counterparts, drawn from three catalogs. This resulted in 7729 \acrlong{COSMOS2015}, 97 \acrlong{i-band}, and 209 \acrlong{IRAC} counterparts to the 
8696 3 GHz sources found within the $1.77\,\mathrm{deg}^2$ subarea of the \gls{COSMOS} field. Spectroscopic redshifts were taken from the \gls{COSMOS} spectroscopic catalog containing 97102 sources {with a counterpart} (M. Salvato, priv. comm.). {In total, 7911 out of the 8035 sources had determined redshifts, of which 2734 had spectroscopic, and 5177 photometric redshifts.} %In total, 7778 out of the 8035 sources had determined redshifts, of which 2740 had spectroscopic, and 5123 photometric redshifts.
To determine the physical properties of the {\gls{AGN} host} galaxies, a three-component \gls{SED}-fitting procedure \citep{Delvecchio17} was applied using all of the available photometry.

As described by \citet{Smolcic:17b}, \gls{AGN} were divided into \gls{HLAGN} and \gls{MLAGN}, which are analogs of high- and low-excitation emission line \gls{AGN}, respectively. As summarized in Fig. 10 in \citet{Smolcic:17b}, \gls{HLAGN} were identified as either X-ray \gls{AGN} \citep[i.e., having X-ray luminosity $L_X>10^{42}\,\mathrm{ergs/s}$,][]{Szokoly04}, mid-infrared \gls{AGN} \citep[using the criteria from][]{Donley12}, or using optical-to-millimeter \gls{SED} fitting \citep{Delvecchio17}. \gls{MLAGN} were identified by requiring red optical colors $M_{NUV}-M_{r^+}>3.5$ and no Herschel-band detections. {The \gls{MLAGN} class contains both \gls{AGN} exhibiting an excess in radio luminosity (described below) and quiescent \gls{AGN}.}

Within this dataset, we use a sample of \gls{RxAGN}.
{The} class of \gls{RxAGN} is identified by their $3\sigma$ radio excess with respect to the value of radio luminosity expected only from the contribution from star formation calculated using the infrared luminosity \citep[see][]{Smolcic:17b, Delhaize17,Delvecchio17}. We separate \gls{RxAGN} into \gls{RxQMLAGN}, \gls{RxSMLAGN}, and \gls{RxHLAGN}, depending on whether they satisfy the \gls{MLAGN} or \gls{HLAGN} criteria.
We have combined {flux densities} at $1.4\GHz$ from the $1.4\GHz$-selected sample of \gls{RxAGN} sources with their corresponding {flux densities} at $3\GHz$. To this dataset, we have added {flux densities} of corresponding sources in the \gls{GMRT} catalogs, matched using \textsc{topcat}.
\FigLz

{In summary, we have selected a sample of \gls{RxAGN} in the \gls{COSMOS} with high completeness for redshifts within $z\in[0,4]$, and $1.4\GHz$ radio luminosities within $\log L_{1.4}\,\mathrm{{[W/Hz]}}\in [24,26]$, as shown in Fig. \ref{fig:lgLz2d}. {In Fig. \ref{fig:Completeness_corr} we show the mean completeness correction for the sample of \acrshort{RxAGN} as a function of redshift. Mean completeness corrections for different redshift bins were computed using the $3\GHz$ flux densities of sources in the \gls{RxAGN} sample. The completeness corrections used were interpolated from the completeness correction table of the \gls{VLA3LP} catalog \citep[see table 2 in][]{Smolcic:17a}. {The mean completeness is $\sim100\%$ up to $z\sim1$ and decreases at higher redshift, but is higher than $\sim75\%$ out to redshift $z=4$.}} In total, our sample comprises 744 \gls{RxAGN}, of which 230 are \gls{RxHLAGN}, 134 \gls{RxQMLAGN} and 380 are \gls{RxSMLAGN}.}
\FigMainSEDs

\section{Methods}
\label{sect:Methods}
We have used the method developed and described in detail by \citet{Tisanic19} to derive survival-analysis based estimates of the average radio \gls{SED}. In Sect. \ref{sect:MethodsSEDoverview} we give a brief overview of the procedure used to construct the average radio SEDs. In \ref{sect:MethodsSEDmodels} we give an overview of models used to describe the radio \gls{SED}.

\subsection{Construction of average radio SEDs}\label{sect:MethodsSEDoverview}
We present the radio \gls{SED} as {flux density} vs. frequency in logarithmic space. To achieve uniform frequency binning, we used equally separated bins in log space of the rest-frame frequencies. For each bin, we computed the mean of the log rest-frame frequency and its standard deviation. {We then combined the normalized {flux densities} and normalized upper limits, {defined as 5 times the local \gls{RMS} value at the position of the source}, of individual galaxies into a single dataset.}  We normalized {flux densities} of both detections and upper limits to a value based
on a linear fit to the $1.4\GHz$ and $3\GHz$ spectra of individual sources, evaluated at the median rest-frame log-frequency of sources in our sample. {If there were no upper limits within the bin, we computed the mean {normalized log-flux density} and its standard deviation for each bin.}
If there were upper limits within a particular bin, we estimated the mean {normalized log-flux density} and its standard deviation from the constructed best-fitting Weibull model of the survival function to the Kaplan-Meier survival function.

Finally, we fit a \gls{BPL} to the {normalized log-flux density}, $\log \tilde F_\nu$, 
\begin{equation}
\log \tilde F_\nu\left(\nu_r\middle| \begin{matrix}\alpha_1\\\alpha_2\\ b\\ \nu_b\end{matrix}\right)=
\begin{cases}
    -\alpha_2 \log \frac{\nu_r}{\nu_n} + C,& \nu_r>\nu_b\\
    -\alpha_1 \log \frac{\nu_r}{\nu_n}+C+(\alpha_1-\alpha_2) \log \frac{\nu_b}{\nu_n},& \nu_r<\nu_b
\end{cases},\label{eq:BPL}
\end{equation}
{{which is} described by the rest-frame frequency, $\nu_r$, normalizing frequency $\nu_n$, the break frequency $\nu_b$, normalization constant $C$, and the spectral indices $\alpha_1$ (below $\nu_b$) and $\alpha_2$ (above $\nu_b$). The fitting was performed by employing the orthogonal distance regression method bounded within the $2\sigma$ confidence interval of parameters, as derived by using the \gls{MCMC} method. }{The normalization constant $C$ is included in the models to ensure statistical validity of our fits and is hereafter implied by the use of ``$\sim$" in the following equations.} We also discuss the \gls{SED} shape by employing models of synchrotron self-absorption and synchrotron aging, as outlined below. 

\subsection{Description of synchrotron spectrum}\label{sect:MethodsSEDmodels}
The spectrum of a synchrotron radio source is influenced by \gls{SSA} and \gls{SA} processes.
Here we briefly describe the influence of both processes on the radio \gls{SED}.

If the synchrotron emission is described as a {power law} with spectral index $\alpha$, the resulting \gls{SSA} \gls{SED} {for a homogeneous source} has the following shape \citep[{constructed using the transfer equation and a power-law \gls{SED} absorption coefficient from}][]{Pacholczyk80}:
\begin{equation}
  \tilde F_\nu\sim \nu^{5/2}\left(1-e^{-(\frac{\nu}{\nu_1})^{-\alpha-5/2}}\right).\label{eq:SSA}
\end{equation}
{This model produces a peak around $\nu=\nu_1$ and a simple power-law \gls{SED} with a spectral index $\alpha$ for $\nu>>\nu_1$.}

If a source is initially described with a spectral index $\alpha$, the spectrum will deviate from a {power law} at later times due to electrons losing energy over time. This produces a spectrum that is steepened by $\Delta\alpha$, {a parameter that varies between single injection and continuous injection models}, at frequencies higher than the break frequency $\nu_b$ \citep[see, e.g.,][]{Condon92}:
\begin{equation}
  \tilde F_\nu\sim \frac{\nu^{-\alpha}}{1+\left(\frac{\nu}{\nu_b}\right)^{\Delta\alpha}}.\label{eq:SA}
\end{equation}

We investigate the combined influence of both effects on the shape of our radio \gls{SED} by introducing a fraction of synchrotron-aged {flux density}, $f$,
\begin{equation}
  \tilde F_\nu\sim f\frac{\nu^{-\alpha_{SA}}}{1+\left(\frac{\nu}{\nu_b}\right)^{\Delta\alpha}}+ (1-f)\nu^{5/2}\left(1-e^{-(\frac{\nu}{\nu_1})^{-\alpha_{SSA}-5/2}}\right).\label{eq:SSA+SA}
\end{equation}
This equation reproduces Eq. \ref{eq:SSA} for $\nu\sim\nu_1<<\nu_b$, and $f\rightarrow 0$, and Eq. \ref{eq:SA} for $\nu_1<<\nu\sim\nu_b$, and $f\rightarrow1$. {The model presented in Eq. \ref{eq:SSA+SA} is chosen so that it can describe a population of sources that have either a synchrotron aged \gls{SED} or a synchrotron self-absorbed \gls{SED}. In this case, the fraction $f$ would represent the fraction of the sample having a synchrotron-aged spectral shape. A further complication to this simple picture would be that a source in the sample has both contributions present, but this would greatly increase the number of free parameters.}

In contrast to simple models discussed in \citet{Tisanic19}, for which the fitting procedure was developed, model of Eq. \ref{eq:SSA+SA} has multiple break frequencies. It is, therefore, a priori unclear which parameters can be fixed to a particular value. {Therefore, to further discuss models of synchrotron self-absorption and aging, we have employed the \acrshort{MBAM} in Sect. \ref{sect:DiscussionCause}. } The method is summarized in Sect. \ref{sect:MethodsMBAM} and is explained in detail in \citet{Transtrum10,Transtrum14}.
\subsection{Manifold Boundary Approximation Method}\label{sect:MethodsMBAM}

{The radio \gls{SED} parameters have been estimated using the orthogonal distance regression constrained to the $2\sigma$ \gls{MCMC}-derived confidence intervals. However, this procedure yielded large estimated parameter errors due to the complex shape of the fitting function. } Moreover, as in \citet{Tisanic19}, the \gls{MCMC} estimates are not simple ellipses, which we had solved by fixing the obvious choice of the greatest uncertainty, the break frequency. Upon applying the same procedure to, {for example}, Eq. \ref{eq:SSA+SA}, we have found that it is not easy to determine the parameters {that have the greatest impact on the derived uncertainties}. We have therefore employed the \gls{MBAM} to determine which of the model parameters contributes to the estimated errors the most \citep[for details, see {for example,}][]{Transtrum10,Transtrum14}.

The \gls{MBAM} method solves the geodesic equation in parameter space to determine the least-precise parameter (or combination of parameters). The method consists of computing the \gls{FIM} from the model parameters (labeled $\theta^\mu$) {following the \citet{Transtrum14} algorithm, described below.}

The method first recognizes that in the space of model residuals of $N_d$ measurements there is an $N_d$ dimensional manifold, $\mathcal{R}$, with \gls{FIM} equaling the Euclidean metric, 
\begin{equation}
  g_{\mathcal{R}}=\sum\limits_{i=1}^{N_d}\mathrm{d}r^i \otimes \mathrm{d}r^i.
\end{equation} Each measurement's residual, $r^i$, is defined as the difference between the data point $y^i$, with uncertainty $\sigma_i$, and the particular model's evaluation $f^i$ as $r^i=(y^i-f^i)/\sigma_i$. The particular model's parameters, which we label $\theta^1,\cdots,\theta^{N_p}$, constitute an $N_p$-dimensional manifold, $\mathcal{M}$. Residuals are therefore an embedding of the model manifold in the residuals' manifold. The metric on $\mathcal{M}$, $g$, is therefore computed using the pullback, $r^*$, as
\begin{equation}
  g=r^* g_{\mathcal{R}}=\left(\sum\limits_{i=1}^{N_d}\partial_\mu r^i\partial_\nu r^i\right)\mathrm{d}\theta^\mu \otimes \mathrm{d}\theta^\nu.
\end{equation}
The geodesic equation in $\mathcal{M}$
\begin{equation}
\frac{\mathrm{d}^2\theta^\mu}{\mathrm{d}\tau^2}+\Gamma^{\mu}{\alpha\beta}\frac{\mathrm{d}\theta^\alpha}{\mathrm{d}\tau}\frac{\mathrm{d}\theta^\beta}{\mathrm{d}\tau}=0,
\end{equation}
is then solved as an initial value problem, where $\Gamma^{\mu}_{\alpha\beta}$ are the Christoffel symbols of the second kind. The initial position is chosen to be the best-fitting parameters $\theta^\mu(\tau=0)=\theta^\mu_{bf}$. The ``velocity'', $\mathrm{d}\theta^\mu/\mathrm{d}\tau(\tau=0)$, is chosen to be the eigenvector of the \gls{FIM} corresponding to the smallest eigenvalue. The components of the eigenvector corresponding to the smallest eigenvalue at a sufficiently large $\tau$ are compared to the values at $\tau=0$.
\FigMBAMa
\section{Results}
\label{sect:Results}
\FigMBAMb
In this section we constrain the shape of the {average} radio \gls{SED} of \gls{RxAGN} and subsets of the \gls{RxAGN} sample. 
In Sect. \ref{sect:ResultsRxAGN}, we derive the \gls{SED} for the \gls{RxAGN} sample, in Sect. \ref{sect:ResultsSubsamples} we describe the \gls{SED}s of subsamples of the \gls{RxAGN} sample and analyze correlations with redshift different \gls{RxAGN} subsamples, source size and further classifications in Sect. \ref{sect:ResultsCorr}.

\subsection{RxAGN SED}
\label{sect:ResultsRxAGN}
{In Fig. \ref{fig:PLSED}, we show the mean {normalized log-flux density} and the corresponding standard deviation of the distribution of {normalized log-flux densities} for each rest-frame frequency bin.} Using a simple \gls{PL} model, the \gls{RxAGN} \gls{SED}, shown in the left panel of Fig. \ref{fig:PLSED}, can be described by a spectral index of \PLSpectralIndex. However, as can be seen from Fig. \ref{fig:PLSED}, this model does not describe the dataset well. Therefore, we {fit} the \gls{BPL} model to account for spectral curvature, as shown in the right panel of Fig. \ref{fig:PLSED}. The spectral indices of this model are \BPLSpectralIndexLower, and \BPLSpectralIndexHigher, while the break frequency, which in the full model was poorly constrained by \BPLBreakFrequency, was fixed to \BPLBreakFixedFrequency to reduce parameter uncertainties, as described in \citet{Tisanic19}. 

{To investigate {how \gls{SA} and \gls{SSA} processes could produce the observed \gls{SED}}, we have {fit} Eq. \ref{eq:SSA+SA} to the \gls{RxAGN} dataset. By blindly applying this model, the \gls{RxAGN} \gls{SED} is described with the following set of parameters, exhibiting varying degrees of uncertainties, \SSASAFraction, \SSASASpectralIndexSA, \SSASASpectralIndexSSA, \SSASABreakFrequencySSA, \SSASABreakFrequencySA, \SSASASpectralBreak. As can be seen in the upper panel of Fig. \ref{fig:MBAM1}, this model describes the data well, but is not {very} informative regarding the constraints of the spectral index and the fraction $f$. We have therefore used the \gls{MBAM} method to reduce the uncertainties of these parameters, {with the resulting \gls{SED} shown in the lower panel of Fig. \ref{fig:MBAM1}.}
We have found that the model is best constrained by setting $\nu_b\rightarrow\infty$, which corresponds to a reduced model described using a synchrotron self-absorbed \gls{SED} with a spectral index labeled $\alpha_{SSA}$, and a power-law \gls{SED} with a spectral index labeled $\alpha_{SA}$ {and without any aging break}. The thus derived model parameters are \SSASASpectralIndexSAMBAM, \SSASASpectralIndexSSAMBAM, \SSASABreakFrequencySSAMBAM and \SSASAFractionMBAM. In Fig. \ref{fig:MBAM1}, we show that this reduced model describes the same data points with a similar \gls{SED} shape as the full model.} 

{Fig. \ref{fig:MBAM2} shows the behavior of the components of the eigenvector corresponding to the smallest \gls{FIM} eigenvalue for the radio \gls{SED} of \gls{RxAGN}. In the best fitting point (middle panel of Fig. \ref{fig:MBAM2}), the eigenvector's components are pointing in a direction that does not purely indicate the most uncertain parameter. However, along a geodesic, for sufficiently large $\tau$, the movement along a geodesic halts in each direction other than $\Delta\alpha$, as indicated by diminishing values of the logarithmic parameter derivatives in the left panel of Fig. \ref{fig:MBAM2}. The eigenvector corresponding to the smallest \gls{FIM} eigenvalue, now computed at the end of the geodesic curve, clearly shows that the least determined parameter is $\Delta\alpha$. Moreover, as indicated by the blue line in the left panel of Fig. \ref{fig:MBAM2}, the geodesic curve significantly varies only in the $\Delta\alpha$ direction, pointing toward $\Delta\alpha\rightarrow\infty$. 
This reduced model also reduces the need for $\nu_b$.}

\subsection{Subsamples}
\label{sect:ResultsSubsamples}
In order to test if the shape of the radio \gls{SED} of \gls{RxAGN} depends on parameters {such as} redshift and radio luminosity, we split the \gls{RxAGN} dataset in bins of redshift and $1.4\GHz$ radio luminosity and {fit} both the \gls{PL} and the \gls{BPL} models to each bin separately. 
The resulting \gls{SED}s are shown in Fig. \ref{fig:split_PL} and the best-fitting \gls{PL} spectral indices are shown for each selected bin in Fig. \ref{fig:split_PL_bins} for the whole \gls{RxAGN} dataset. {The bins, whose boundaries are outlined by black dashed lines in Fig. \ref{fig:split_PL_bins}, split the \gls{RxAGN} sample into four bins and additionally include two bins, for lower and higher $1.4\GHz$ radio luminosities. All bins have a mean completeness correction higher than $75\%$. } 

{We have further used source sizes of \gls{RxAGN} \citep{Bondi18} to investigate {whether there is a correlation of spectral indices with the source size, $D$, or the type of the source (\gls{RxHLAGN}, \gls{RxSMLAGN}, \gls{RxQMLAGN}) in different bins. To this end, we further split the dataset in mutually independent subsets of \gls{RxHLAGN}, \gls{RxQMLAGN} and \gls{RxSMLAGN} (see Sect. \ref{sect:DataSample} for details). In Fig. \ref{fig:corrplot} we show the dependence of the \gls{BPL} spectral indices on source size for subsets of the \gls{RxAGN} dataset binned in redshift and $1.4\GHz$ radio luminosity.} }
We show the determined spectral indices for various bins in Table \ref{tab:BPLresults}. 

In order to investigate the possibility that there might be sources in our sample that are severely influenced by \gls{SSA}, we split our sample by selecting all sources having source sizes less than $1\kpc$. This size is expected for gigahertz-peaked sources \citep{ODea98,Collier2018High-resolutionSources}. From the turnover frequency-linear size relation \citep{Orienti14} we expect that this requirement should {select} sources with turnover frequencies above $1\GHz$. The constraint on source size less than $1\kpc$ produces a sample of 25 sources, for which we construct a \gls{PL} and \gls{BPL} \gls{SED}.

The \gls{PL} model yields an \gls{SED} with a spectral index of \flatPL, while the \gls{BPL} model describes the spectrum with a spectral index of \flatBPLlow below \flatBPLFrequency and {\flatBPLhigh above $\nu_b$}. The resulting \gls{SED} is shown in Fig. \ref{fig:inv}. 
\subsection{Significance of correlations}
\label{sect:ResultsCorr}
{In Fig. \ref{fig:corrplot2}, we show the components of the correlation matrix with given p-values of correlation coefficients. If we adopt a p-value cut, $P<0.1$, this correlation matrix shows the following.}
{Firstly, there is a positive correlation {of the \gls{PL} spectral index $\alpha$} with both {the} redshift and radio luminosity. Secondly, there is a positive correlation {of the \gls{PL} spectral index $\alpha$} with both {the} redshift and radio luminosity. And thirdly, there is a positive correlation of the \gls{BPL} spectral index $\alpha_2$ with source size, radio luminosity and the \gls{PL} spectral index $\alpha$.}

{We have further tested the level of significance of the various correlations presented in the correlation matrix. To this end, we have computed the p-values of parameters in linear models for all subsets of parameters using the \gls{ANOVA} method, as implemented in the \textsc{statsmodels} package \citep{seabold-proc-scipy-2010}. The goal of this method was to minimize overfitting the spectral indices, while retaining the ability to discern models by how well they explain the variability between the subsets of \gls{RxHLAGN}, \gls{RxQMLAGN} and \gls{RxSMLAGN}. The \gls{ANOVA} results are shown in Table \ref{tab:ANOVA} and are based on spectral indices presented in Table \ref{tab:BPLresults} and visualized in Fig. \ref{fig:violinplot}. We find that the \gls{PL} spectral index and the \gls{BPL} spectral index $\alpha_2$ can both be described by a model containing only redshift and source size with a significance value of $P<0.01$ . The parameters of the corresponding linear models are}
\begin{align*}
\alpha &= 0.39\pm0.07+ (0.17\pm0.03)z\\
\alpha_2 &= 0.6\pm0.2+ (0.27\pm0.09)\log D\mathrm{{[kpc]}}+ (0.15\pm0.05)z,
\end{align*}
which is in line with correlations from the correlation matrix.

{We investigate weaker variations that might be present between the spectral index subsets by employing a p-value cut of $P<0.1$, the least stringent of the commonly used p-value cuts. We find that only the \gls{BPL} spectral index $\alpha_1$ can be described using source type and source size.
The parameters of the corresponding linear models are}
\begin{align*}
\alpha_1 &= (0.0\pm0.2)+ (0.3\pm0.2)\log D\mathrm{{[kpc]}}\\
\alpha_1&=\begin{cases}0.0\pm0.1+(0.3\pm0.1)\log\mathrm{{[kpc]}}-0.2\pm0.1,\, \mathrm{RxSMLAGN}\\
0.0\pm0.1+(0.3\pm0.1)\log D\mathrm{{[kpc]}},\,\mathrm{for\,others}
\end{cases}.
\end{align*}

\FigSplitSEDs
\FigSplitBins
\TabIndices
\FigCorrFirst
\section{Discussion}
\label{sect:Discussion}
In this section we explore the properties of the derived radio \gls{SED} of \gls{RxAGN}. In Sect. \ref{sect:DiscussionCause}, we investigate the cause for the derived \gls{SED} shape. In Sect. \ref{sect:DiscussionTrends} we analyze possible correlations with redshift of different \gls{RxAGN} subsamples, source size and further classifications, while in Sect. \ref{sect:DiscussionFlat} we discuss the \gls{SED} for the subsample of flat-spectrum sources.
\subsection{Possible cause of the SED shape}

\label{sect:DiscussionCause}
It is clear from Fig. \ref{fig:PLSED} that a simple \gls{PL} model does not describe the \gls{SED} well. Using the \gls{BPL} model yields an \gls{SED} shape with a spectral index difference $\Delta\alpha=\alpha_2-\alpha_1=0.88\pm0.05$. This spectral index difference is higher than expected from the simple Kardashev-Perola model of synchrotron aging \citep[$\Delta\alpha=0.5$,][]{Kardashev62,Kardashev62Cyg}. This may indicate that the \gls{SED}s of individual sources may be better described by either more complex models of synchrotron aging \citep{JP73,Tribble2014RadioField} or the \gls{SED} shapes are affected by processes of synchrotron self-absorption or free-free absorption \citep{Menon1983RedshiftQuasars,Tingay,Kameno}.

The \gls{SED} shape derived by the \gls{MBAM} method shows a { ``bump'' at $\sim 3\GHz$ and a low frequency steepening around $0.6-1\GHz$}. 
For sources detected at both GMRT and VLA frequencies, there seems to be a deficit in flux computed at $610\MHz$ based on the \gls{VLA} fluxes. This correction is appropriate for some of the sources, while blindly applying a $20\%$ correction of the total {flux density} {introduced overcorrections} in others. The {flux density} offset is different when computed based on the $1.4\GHz$ catalog and when computed using the $3\GHz$ catalog, indicating complications due to complex spectral shapes. Since this frequency range is dominated by the $610\MHz$ GMRT catalog fluxes, to account for possible {biases in the fluxes at this frequency}, we have performed {again} the \gls{SED} fitting procedure without the $610\MHz$ catalog. We find a similar {steepening} in the \gls{SED} when using only the $325\MHz$, $1.4\GHz$ and $3\GHz$ data, indicating that the effect may be due to the {intrinsic} shape of the \gls{SED} and not due to systematic effects of a particular catalog.
Alternatively, this result could explain the findings of \citet{CalistroRivera17}, since the observed feature in our \gls{SED} effectively produces a steeper spectrum below $1\GHz$ with a flattening around $1\GHz$. Moreover, the fraction of \gls{SSA} emission ($1-f=0.2\pm0.03$, {obtained using Eq. \ref{eq:SSA+SA} and the \gls{MBAM} procedure}) described by a steep \gls{SSA} spectral index is in line with \citet{Kapahi1981WesterborkSurvey.} and \citet{Peacock1982BrightTelescope}, suggesting the presence of gigahertz-peaked or inverted spectrum sources in our sample.

{As an alternative, we have {fit} the free-free absorption models both in the form of a foreground-screen and as a mixed model, as described in \citet{Tisanic19}. The foreground-screen model has a spectral index of $\alpha=0.8\pm0.1$ with {an} optical depth at $1\GHz$ of $\tau_1=0.2\pm0.1$, while the mixed model has a spectral index of $\alpha=0.9\pm0.1$ and an optical depth at $1\GHz$ of $\tau_1=1.3\pm0.5$.} 
\TabANOVA
\subsection{SED shape dependence on various observables}
\label{sect:DiscussionTrends}

 The \gls{BPL} spectral indices {are rising} with source size, which would imply that the {radio} \gls{SED}s of larger sources are more influenced by larger-scale features (jets). {\citet{Blundell1999TheSamples} argue that sources at higher redshift are on average younger than sources at lower redshift and thus have interactions of their jets occurring closer to the host galaxy. This could explain the trend of the rising  \gls{BPL} spectral index $\alpha_2$ with both {the} redshift and source size. \citet{Ker2012NewSamples} find that the strongest correlation is between spectral index and source size, with redshift correlation becoming more important around $\sim 1\GHz$. This might explain the \gls{ANOVA} giving higher importance to redshift dependence for the higher frequency \gls{BPL} spectral index ($\alpha_2$) than for $\alpha_1$.} Furthermore, only the spectral index below $4\GHz$ shows (weak) correlation with the type of source. {The \gls{ANOVA} test suggests a $2\sigma$ lower spectral index  $\alpha_1$ for \gls{RxSMLAGN} than for other subsets, as shown in Fig. \ref{fig:violinplot} } 

For a $10\kpc$ sized source which is close to the values of $\log D\,\mathrm{{[kpc]}}$ for bins listed in Table \ref{tab:BPLresults} these correlations would imply a broken power-law {radio} \gls{SED} described by a spectral index of $0.9\pm0.2$ above $4\GHz$ and a spectral index $(0-0.3)\pm0.1$ below $4\GHz$, with the higher value estimated for \gls{RxHLAGN} and the lower value estimated for \gls{RxSMLAGN}. As a consistency check, if we approximate our \gls{RxAGN} sample as having $z\sim 2$, we are in agreement with our parameters of $\alpha_1$ and $\alpha_2$ for the average radio \gls{SED} of \gls{RxAGN}, {reported in Sect. \ref{sect:Results}.}

The positive correlation of spectral indices with source size may be explained by the size-{turnover frequency} relation for gigahertz-peaked sources \citep{Orienti14}. 
Smaller source size ($<1\kpc$) implies a high turnover frequency, thereby producing smaller $\alpha_1$ values in an averaged sample. The higher-frequency spectral index could also be affected by this relation, since the spectral peak could be broadened, producing a flatter {radio} \gls{SED} \citep{Odea91}. The redshift dependence could be related to the difference in luminosity functions of flat and steep spectrum sources \citep{Jarvis02}. 
\subsection{Flat spectrum sources}
\label{sect:DiscussionFlat}
 Both the \gls{PL} and the \gls{BPL} spectral indices classify our sample below $1\kpc$ as having a flat spectrum by the usually accepted $\alpha<0.5$ criterion \citep[see, e.g.,][]{deZotti10}. Since the break frequency is larger than $1\GHz$, the $<1\kpc$ sample does not show a significant contribution of compact steep-spectrum sources \citep[a spectral index of $\sim 0.75$ above a turnover frequency {of} $0.5\GHz$, ][]{Kapahi1981WesterborkSurvey.,Peacock1982BrightTelescope, ODea98, deZotti10}.

We find a spectral index difference of $\Delta\alpha=\alpha_2-\alpha_1=0.45\pm 0.1$, which is less than the theoretical limit imposed by pure synchrotron self-absorption \citep[][]{Mhaskey19}. {This indicates that other absorption processes may also influence the flat spectrum radio \gls{SED}, {such as} free-free absorption \citep{Kameno, Tingay}. 
We have performed a fit of the \gls{SSA} model on the $<1\kpc$ sample, yielding a flat spectral index of \flatSSA above a break frequency of \flatSSAFreq. 
A broader rest-frame frequency range is needed to determine if this is due to a single \gls{SSA} component, or a result of multiple break frequencies \citep{Kellermann1969TheSources,Cotton80}.}
{As an alternative, we have {fit} the free-free absorption models both in the form of a foreground-screen, wherein absorption occurs in front of the source, and as a mixed model, wherein absorption occurs within the source, as described in \citet{Tisanic19}. The foreground-screen model has a spectral index of $\alpha=0.6\pm0.1$ with {an} optical depth at $1\GHz$ of $\tau_1=0.4\pm0.1$, while the mixed model has a spectral index of $\alpha=0.6\pm0.1$ and an optical depth at $1\GHz$ of $\tau_1=1.2\pm0.4$. Both models point to a similar spectral index and differ in optical depth at $1\GHz$.}
\section{Summary}
We have constrained the {average} radio \gls{SED} for a sample of \gls{AGN} in the \gls{COSMOS} field using available \gls{VLA} and \gls{GMRT} data in the rest-frame frequency range from $\sim 0.3\GHz$ to $\sim 10\GHz$. { The radio-excess \gls{AGN} (\gls{RxAGN}) sample contained sources which exhibit a $3\sigma$ excess of $1.4\GHz$ radio luminosity, compared to that expected solely from star-forming processes occurring within the \gls{AGN} host galaxies. The \gls{RxAGN} sample is relatively complete {($75\%$)} in redshift out to $z\sim 4$ and in radio luminosity range from $10^{24}\,\mathrm{W/Hz}$ to $10^{26}\,\mathrm{W/Hz}$.}

We find that if we fit the radio \gls{SED} for this sample with a single power-law model, the resulting spectral index is \PLSpectralIndex. However, such a model does not capture all the features of {the \gls{RxAGN} radio \gls{SED} which can be better described} by a broken {power law} with a spectral index of \BPLSpectralIndexLower, below $4\GHz$, and \BPLSpectralIndexHigher, above $4\GHz$. {The derived {radio} \gls{SED} can be even better described by models involving both \acrlong{SSA} and \acrlong{SA} processes.}

{By binning in both the $1.4\GHz$ radio luminosity and redshift, we have found (at $P<0.01$ significance) that the single power-law spectral index is positively correlated with redshift and that the broken power-law spectral index above \BPLBreakFixedFrequency is positively correlated with both {the} redshift and source size.
With a somewhat lesser significance ($P<0.05$), we find that also the $\gls{BPL}$ spectral index below \BPLBreakFixedFrequency is positively correlated with source size. For a subsample of sources having sizes less than $1\kpc$, we find a flat spectrum {radio} \gls{SED} described by a single power-law spectral index of \flatPL. Using the broken power-law model we find a flat spectrum {radio} \gls{SED} with a spectral index of \flatBPLlow (\flatBPLhigh) below (above) \flatBPLFrequency.}

{Our derived average radio \gls{SED}s of \gls{RxAGN} span a frequency range from $300\MHz$ to $10\GHz$ and thus cover a significant portion of the expected \gls{SKA} frequency window. The spectral indices we find in the frequency range from $300\MHz$ to $4\GHz$ are significantly lower than the often assumed values in large surveys around $1\GHz$, indicating a greater impact of complex radio SED shapes on the derived properties of galaxies. Our radio SEDs offer a way to better understand future observations in the $300\MHz-10\GHz$ regime by going beyond the often assumed simple power-law radio SED shape.  }
\section*{Acknowledgements}
This research was funded by the European Union's Seventh Framework program under grant agreement 337595 (ERC Starting Grant, `CoSMass'). JD acknowledges financial assistance from the South African Radio Astronomical Observatory (SARAO; www.ska.ac.za).

\bibliography{references}

\end{document}